\documentclass[fleqn,10pt]{wlscirep}

\usepackage{xcolor}					  %
\definecolor{green}{RGB}{52,200,50}    %
\title{The role of initial speed in projectile impacts into light granular media}
\author[1,2*]{Kai Huang}
\author[3]{Dariel Hernadez Delfin}
\author[2]{Felix Rech}
\author[2]{Valentin Dichtl}
\author[3,+]{Ra\'ul Cruz Hidalgo}
\affil[1]{Division of Natural and Applied Sciences, Duke Kunshan University, 215306 Kunshan, Jiangsu, China}
\affil[2]{Experimentalphysik V, Universit\"at Bayreuth, 95440 Bayreuth, Germany}
\affil[3]{Department of Physics and Applied Mathematics, University of Navarra, 31009 Pamplona, Spain}

\affil[*]{kh380@duke.edu}
\affil[+]{raulcruz@unav.es}
\begin{abstract}

Projectile impact into a light granular material composed of expanded polypropylene (EPP) particles is investigated systematically with various impact velocities. Experimentally, the trajectory of an intruder moving inside the granular material is monitored with a recently developed non-invasive microwave radar system. Numerically, discrete element simulations together with coarse-graining techniques are employed to address both dynamics of the intruder and response of the granular bed. Our experimental and numerical results of the intruder dynamics agree with each other quantitatively and are in congruent with existing phenomenological model on granular drag. Stepping further, we explore the `microscopic' origin of granular drag through characterizing the response of granular bed,  including density, velocity and kinetic stress fields at the mean-field level. In addition, we find that the dynamics of cavity collapse behind the intruder behaves qualitatively different with different impact velocities. Moreover, the kinetic pressure ahead of the intruder decays exponentially in the co-moving system of the intruder. Its scaling gives rise to a characteristic length scale, which is in the order of intruder size. This finding is in perfect agreement with the long-scale inertial dissipation type that we find in all cases.

\end{abstract}
\begin{document}

\flushbottom
\maketitle

\thispagestyle{empty}

\section*{Introduction}

We feel the drag force from air while walking, jogging or riding a vehicle. We also feel the drag force from sand while walking on the beach. The question is to which extent the drag force laws known for Newtonian fluid can be applied to granular materials, such as sand, powders and grains. Such an extension can not only shed light on a fundamental understanding of drag force in non-Newtonian fluids \cite{Chhabra2006, Katsuragi2016, Meer2017}, 
but also on widespread applications of granular materials including pile driving in soil mechanics, impact cratering in geo- and astrophysics, and last but not least, robotic locomotion \cite{Ruiz2013, Hosoi2015}. 
 
Drag induced by projectile penetration is an interesting topic with a long history, particularly in association with ballistic impact \cite{Robins1742,Poncelet1829}. In the past decades, the interest has been broadened in connection with crater formation \cite{Uehara2003,Bruyn2004,Katsuragi2007,Brzinski2013}, partly motivated by space exploration \cite{Colaprete2010,Shinbrot2017} as well as deciphering the 
history of planet formation \cite{Ruiz2013}.  Moreover, it has been extended to the influence of air cavity collapsing as well as granular jet formation \cite{Royer2005}, the projectile shape \cite{Brzinski2013,Askari2016,Kang2018} and the role of gravity on the final penetration depth and time \cite{Altshuler2014}.  As an extreme case, infinite penetration of a projectile into light-weighted EPS (expanded polystyrene) particles was observed, when the weight of the projectile exceeds a critical value \cite{Pacheco2010}.

Based on experimental and numerical investigations, an empirical law based on Poncelet model \cite{Tsimring2005,Katsuragi2016} is typically used to describe the penetration curve of a projectile. It includes a `hydrostatic' term dependent on the penetration depth $z$ \cite{Katsuragi2007} and  a drag term $\propto v^2$ with $v$ the instantaneous speed of the projectile with respect to the granular medium. The {\it inertial} nature of this drag force  has been proposed in the past  \cite{Katsuragi2007,Pacheco2011,Lopez2017}. Nevertheless, that is not the complete story, as recent investigations also reveal a variation of the scaling with $v$ for vibrofluidized granular materials \cite{Seguin2017}. Moreover, collisional damping $\propto v^2$ can also be understood, as a `viscous' scale damping process, governed by a speed dependent collisional rate \cite{Crassous2007,Seguin2009} and the fluctuations of force chains with a stochastic nature \cite{Clark2012,Takehara2014}. In short, a first-principle theory for describing granular drag is still lacking, partly owing to the challenges in predicting the rheological response of granular media, particularly on the brink of solid-liquid-like transition \cite{Jaeger1996,Duran2000}.

From the perspective of applications, EPS or EPP particles are widely used for construction, packaging, damping, as well as raw materials for molded forms, due to its light yet rigid, and good thermal isolation properties \cite{Basf}. The drop tower facility in Bremen \cite{Zarm}, for instance, uses EPS particles regularly as damping materials to capture drop capsules that weight typically over $400$~kg. As such, a better understanding of granular drag, particularly the physical mechanisms behind the empirical laws, is desirable.

\begin{figure}
\includegraphics[width = 0.45\columnwidth]{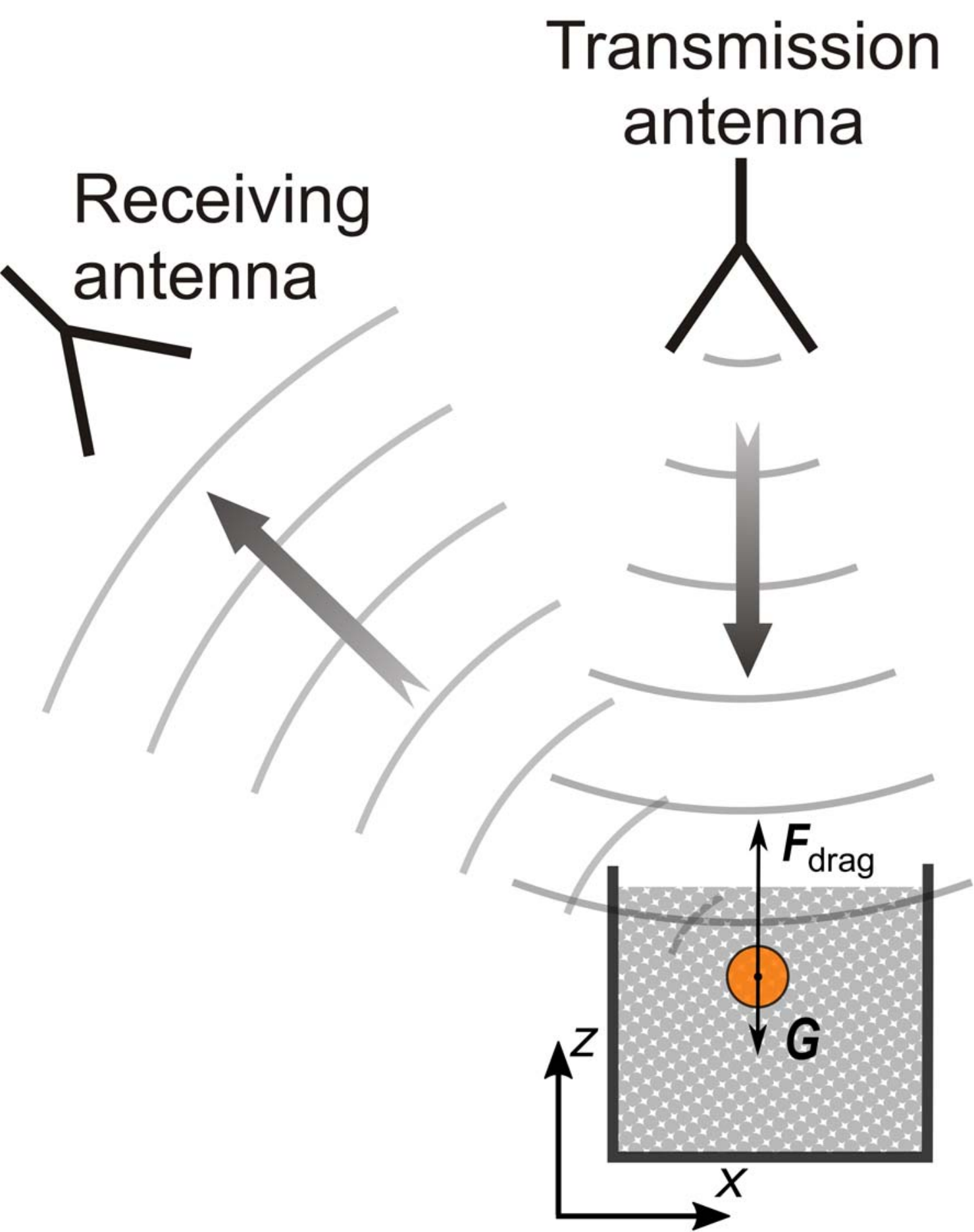}
\caption{\label{fig:setup}(color online) Schematic of the radar tracking set-up viewing from the side. The radar system compares the relative distance between the transmission and receiving antennae (only one sketched here) to obtain the projectile trajectory inside the EPP particles, which are transparent to EM waves.}
\end{figure}

Here, we employ a recently developed microwave radar system that `sees' through granular particles and DEM (discrete element method) simulations \cite{Altshuler2014} to investigate the influence of the impact velocity on the drag force induced by light weighted EPP particles (Neopolen, P9255) with a mean diameter of $3$~mm on a spherical projectile with a diameter $D=1$~cm. Experimentally, the position of the projectile is determined from the phase shift of $10$~GHz electromagnetic waves travling from the transmission to the receiving antennae with a triangulation process \cite{Ott2017}, as sketched in Fig.\,\ref{fig:setup}. The experimental conditions, including the dimensions of the intruder, granular particles, the container, as well as the material properties of the particles, are followed in the numerical simulations, which allows us to explore the origin of granular drag through characterizing the response of granular particles at the mean-field level. The results suggest that, despite the order-of-magnitude difference in the kinetic stress and granular energy levels, the profiles of scaled kinetic stress viewed from the perspective of the projectile is surprisingly independent on the initial speed of the intruder $v_{\rm i}(0)$. In another word, the kinetics of the projectile inside of a granular medium is pre-determined by the initial speed of the intruder $v_{\rm i}(0)$.

%%%%%%%%%%%%%%%%%%%%%%%%%%%%%%%%%%%%%%%%%%%%%%%%%%%%%%%%%%%%
%                  Results and Discussions
%%%%%%%%%%%%%%%%%%%%%%%%%%%%%%%%%%%%%%%%%%%%%%%%%%%%%%%%%%%%
\section*{Intruder dynamics}

\begin{figure*}%[b]
\begin{minipage}[t]{0.5\textwidth}
{\large (a)} \\
\includegraphics[width = 0.9\textwidth]{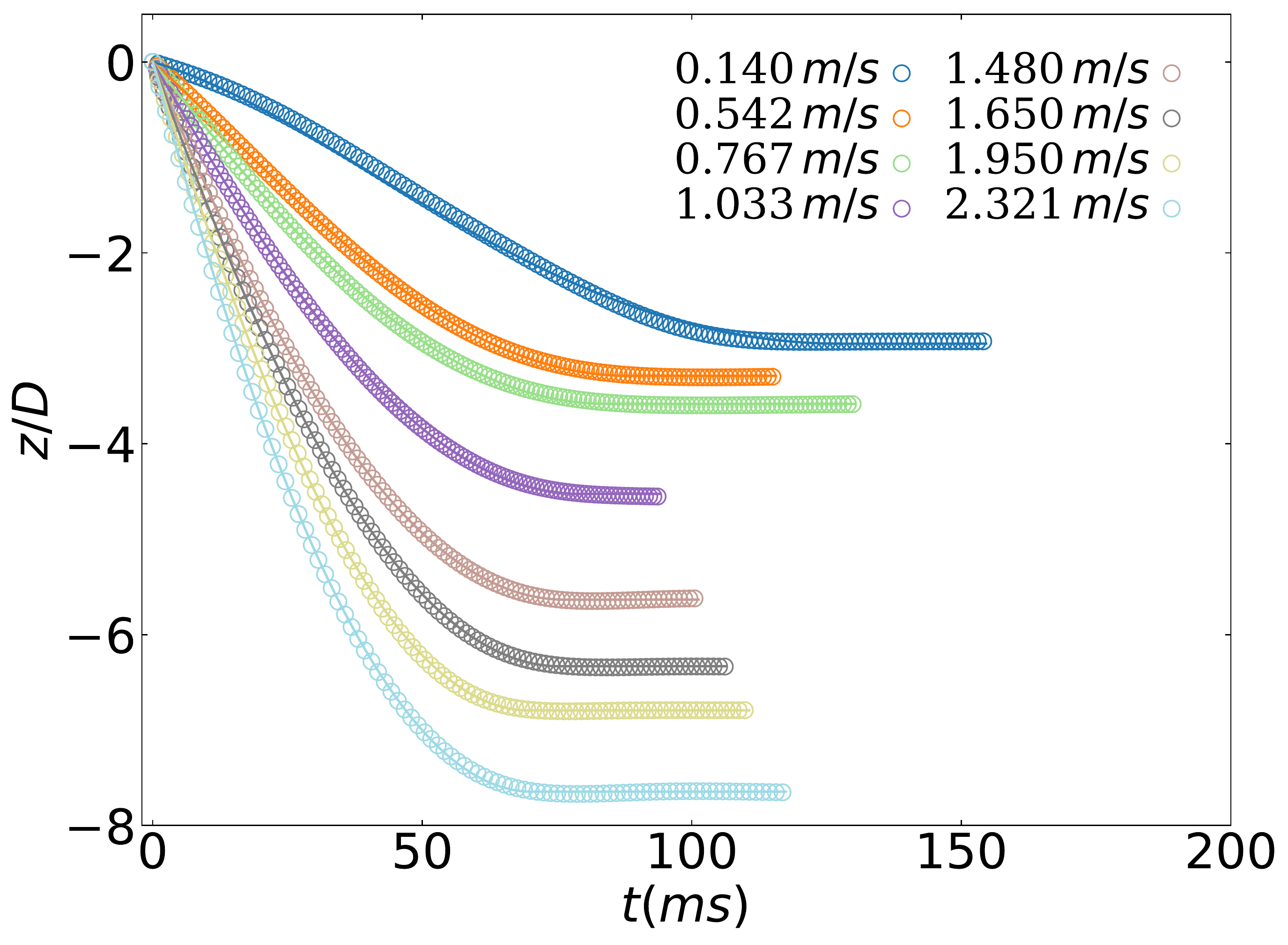} 
\end{minipage}
\begin{minipage}[t]{0.5\textwidth}
{\large (b)} \\
\includegraphics[width = 0.9\textwidth]{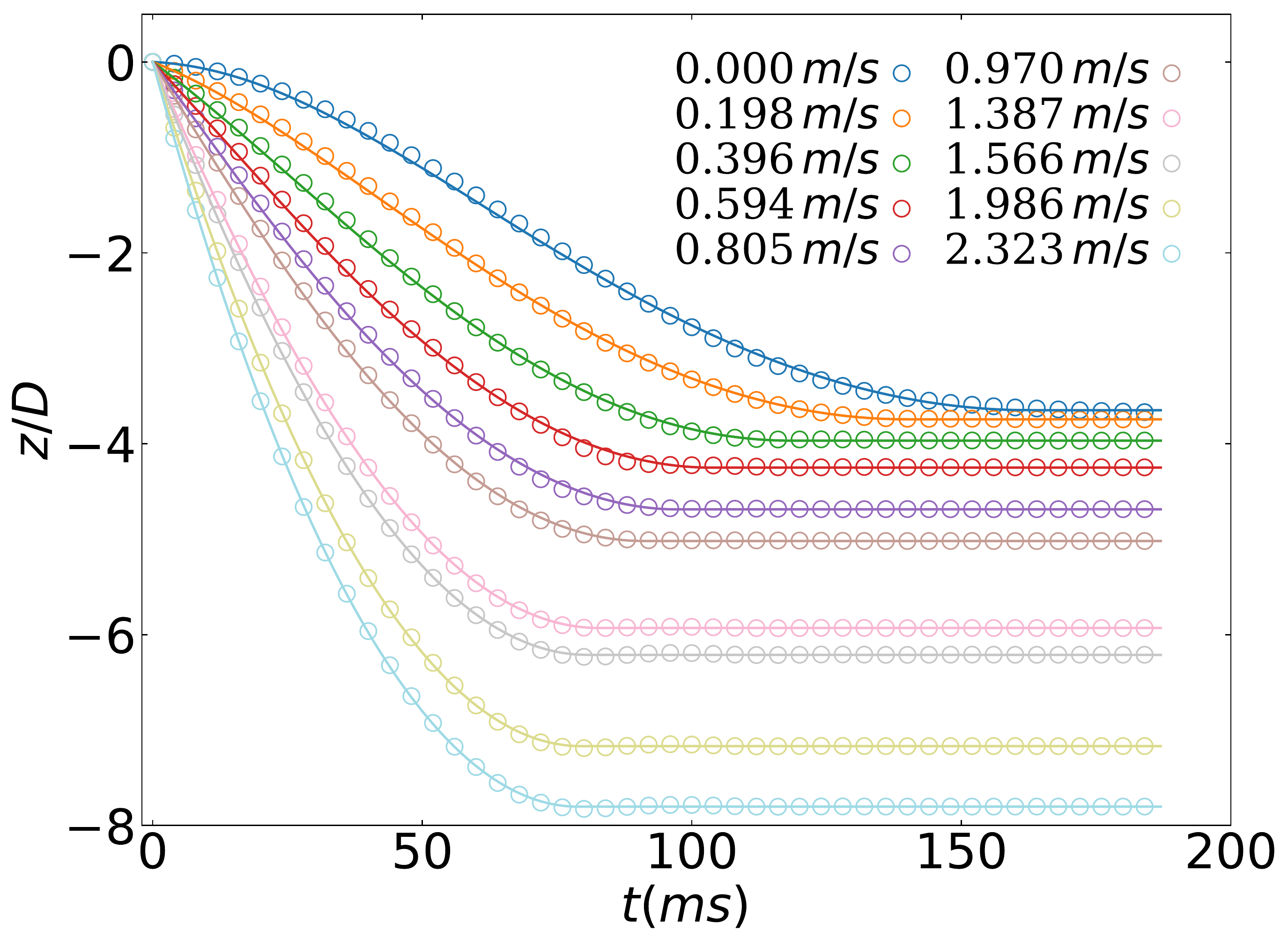} 
\end{minipage}
\caption{Sample penetration curve z(t); experiment a)  and simulation  b). In both cases, the data is compared with the empirical model Eq.(\ref{eq:motion}).}
\label{fig:zt}
\end{figure*}

\begin{figure*}%[b]
\begin{minipage}[t]{0.5\textwidth}
{\large (a)} \\
\includegraphics[width = 0.9\textwidth]{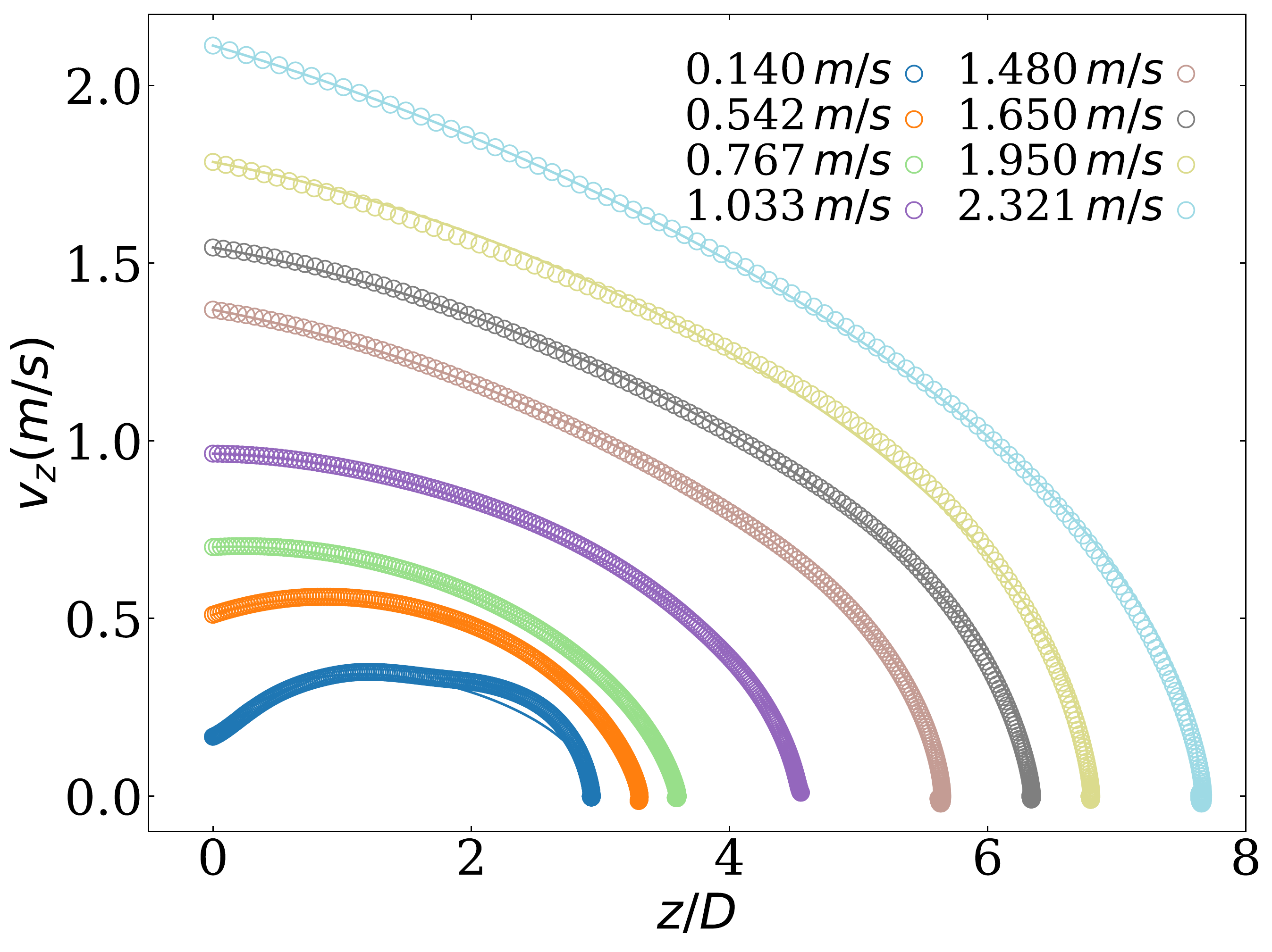} 
\end{minipage}
\begin{minipage}[t]{0.5\textwidth}
{\large (b)} \\
\includegraphics[width = 0.9\textwidth]{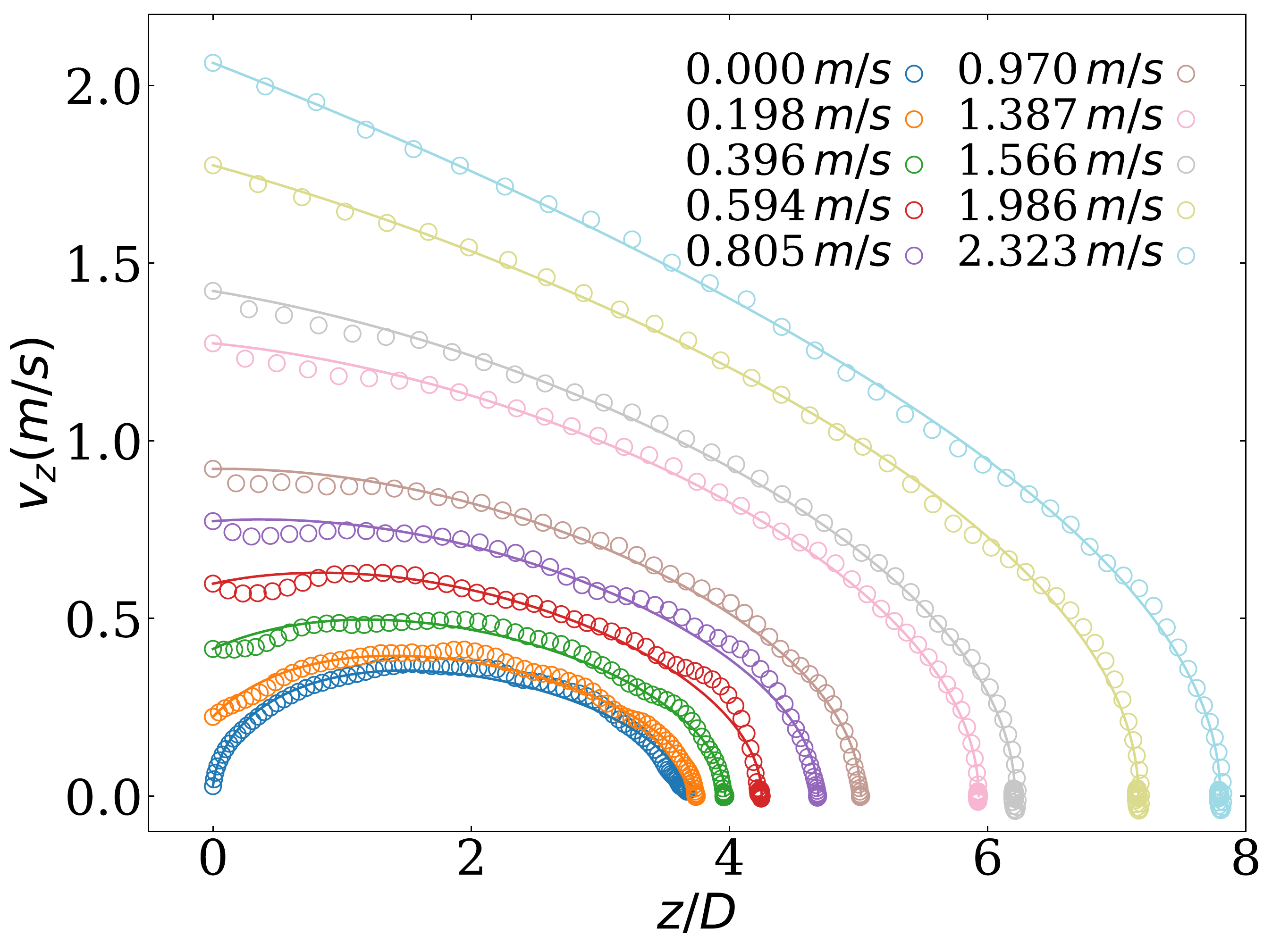} 
\end{minipage}
\caption{Intruder velocity $v_z(t)$; experiment a)  and simulation  b). In both cases, the data is compared with the empirical model
Eq.(\ref{eq:motion}).}
\label{fig:Vzt}
\end{figure*}

We perform a systematic study of the intruder penetration dynamics at various initial impact speed ($v_{\rm i}=[0-2.3]$~m/s). Figs.\ref{fig:zt} and \ref{fig:Vzt} show the vertical position $z(t)$ and speed $v_{\rm i}(t)$ of the intruder, obtained experimentally and numerically, respectively. For the experimental cases, the initial speed is estimated from a parabolic fit of the intruder trajectory before impact together with the initial falling height $H$. As noted, in all cases the intruder dynamics predicted numerically is in very good quantitative agreement with the experimental results and in line with previous findings \cite{Ciamarra2004,Pacheco2011,Katsuragi2013,Xu2014}. 

In general, the intruder dynamics is predominately conditioned by the initial conditions. For all cases, the intruder penetrates into the medium, slowly decelerates until stops at a certain depth. For low impact speeds, there exists an initial accelerating phase as gravity exceeds granular drag at impact, instead of continuous deceleration as in the cases with high impact velocities. Quantitatively, the intruder movement along the gravity direction can be described using its translational equation of motion.
\begin{equation} 
\label{eq:motion}
m\frac{dv_{\rm i}}{dt} = mg-\gamma v_{\rm i}^2 -\kappa \lambda \left(1-e^{-z/\lambda}\right),
\end{equation}
\noindent 
which includes gravity $mg$, drag force and depth-dependent frictional terms with $\gamma$, $\kappa$ and $\lambda$ phenomenological parameters. Following a previous investigation \cite{Pacheco2011}, the characteristic lenght scale $\lambda$ is chosen to be the radius of the container $15\, cm$, which is sufficiently large to avoid boundary effect. The frictional term is linear and scale-free for small penetration depths and it saturates to $\kappa z$ for large scales \cite{Pacheco2011,Katsuragi2013}. In the past, several experimental works showed that the quadratic dependence in speed is sufficient to describe the mechanical energy dissipation of the intruder \cite{Pacheco2011,Katsuragi2013}. The energy dissipation from granular drag resembles an object moving in a fluid at high Reynolds number. The numerical solutions of Eq.(\ref{eq:motion}) is also included in Fig.\ref{fig:zt} and Fig.\ref{fig:Vzt}, for comparison in each case.

\begin{figure}%[b]
\includegraphics[width = 0.48\textwidth]{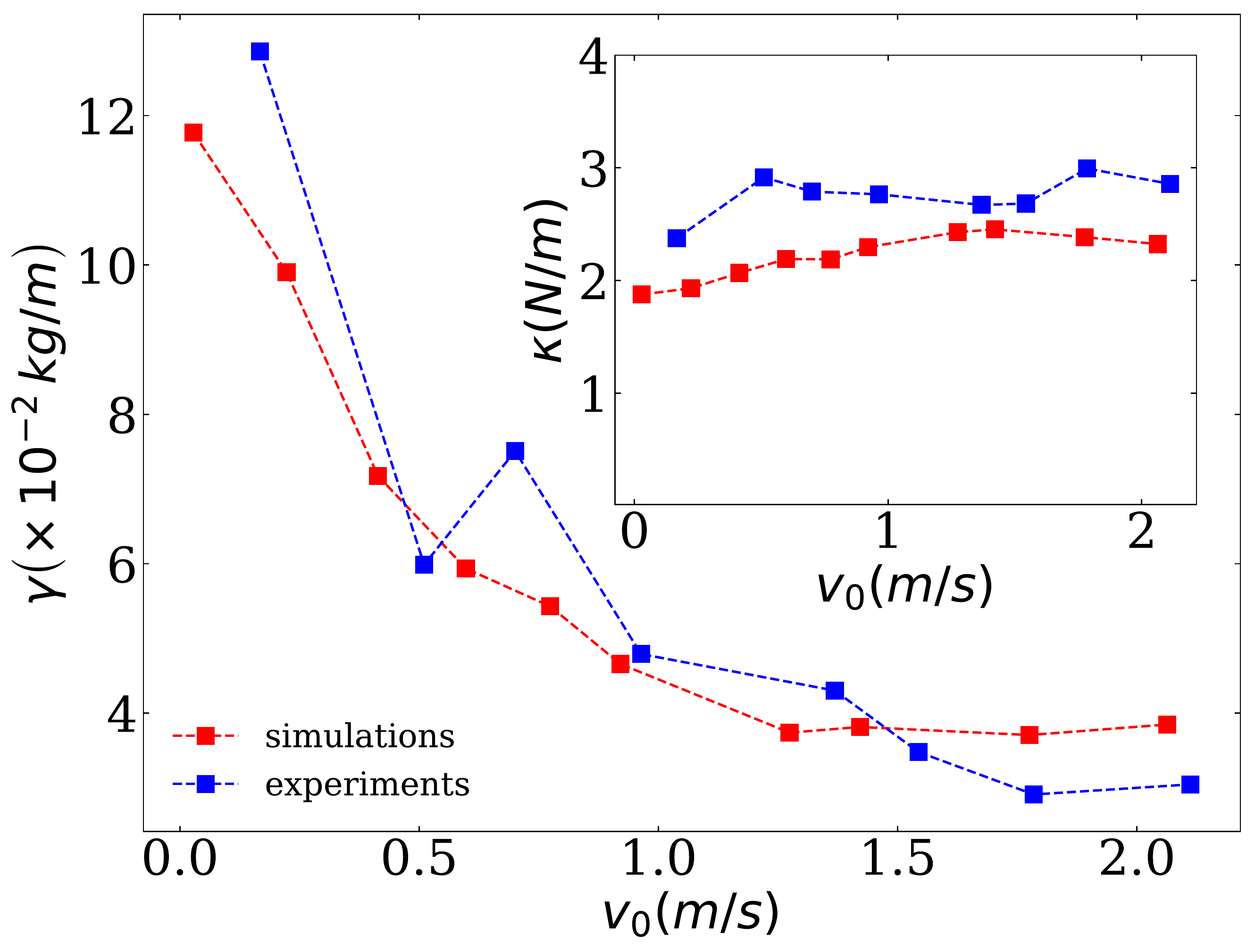} 
\caption{The parameters of the empirical model $\gamma$ and $\kappa$ (see inset), 
obtained for different initial velocities of the intruder $v(0)$. In both graphs,
 experimental and numerical values are shown for comparison.}
\label{fig:gamma_kappa}
\end{figure}

Nevertheless, we find the coefficient $\gamma$ (see Fig.\ref{fig:gamma_kappa}) is not a material constant, but decays with increasing $v_{\rm i}(0)$ until it saturates at high impact speed. It suggests different mechanical response at low $v_{\rm i}$, which arises presumably from different perturbation distance in comparison to impact at high $v_{\rm i}$. Assuming that the coefficient $\gamma$ can be estimated from the momentum transfer, $\gamma v_{\rm i}^2 \propto A~\rho_{\rm p} \varphi ~v_{\rm i}^2$ with $A$ projected projectile area \cite{Katsuragi2013}, there outcomes would suggest that the initial impact reduce notably the packing fraction $\varphi$ of the granular bed. Indeed, this trend was found earlier, examining intruders in fluidized beds, using a up-flow air supply \cite{Brzinski2013}. In our experiment, however,  we do not detect significant changes in volume fraction and, accordingly, our outcomes can not be explained following the same arguments. Moreover, we find the dependency of $\kappa$ on the initial velocity is notably weaker \cite{Brzinski2013} (see inset in Fig.\ref{fig:gamma_kappa}). It is worth mention, that the intruder's trajectory Eq.~\ref{eq:motion}
has been also described, introducing a depth dependent $\gamma$ in terms of its initial kinetic energy \cite{Clark_2013,Bester2017}.

\begin{figure}
{\large (a)} \\
\includegraphics[width = 0.45\textwidth]{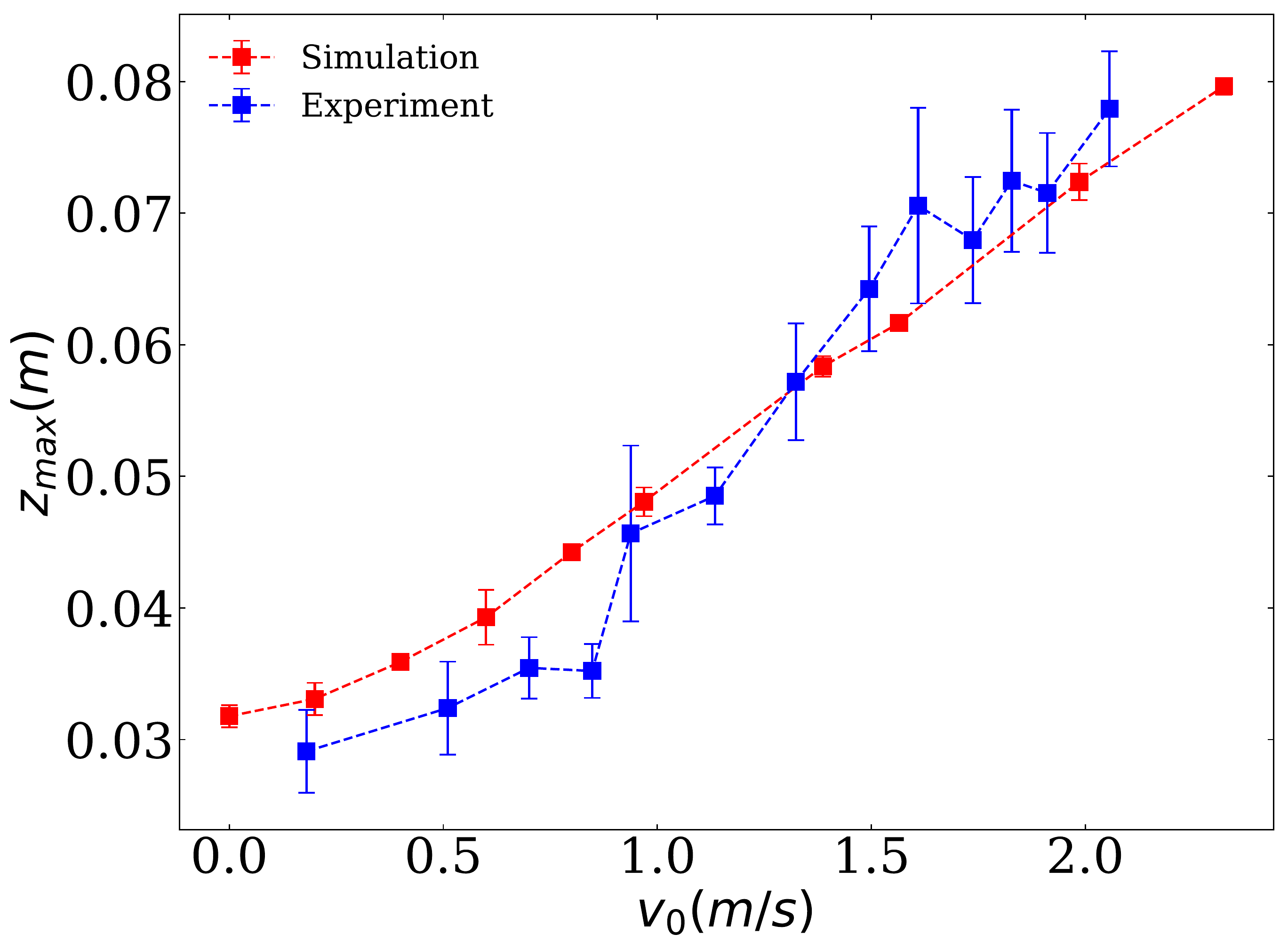} \\
{\large (b)} \\
\includegraphics[width = 0.45\textwidth]{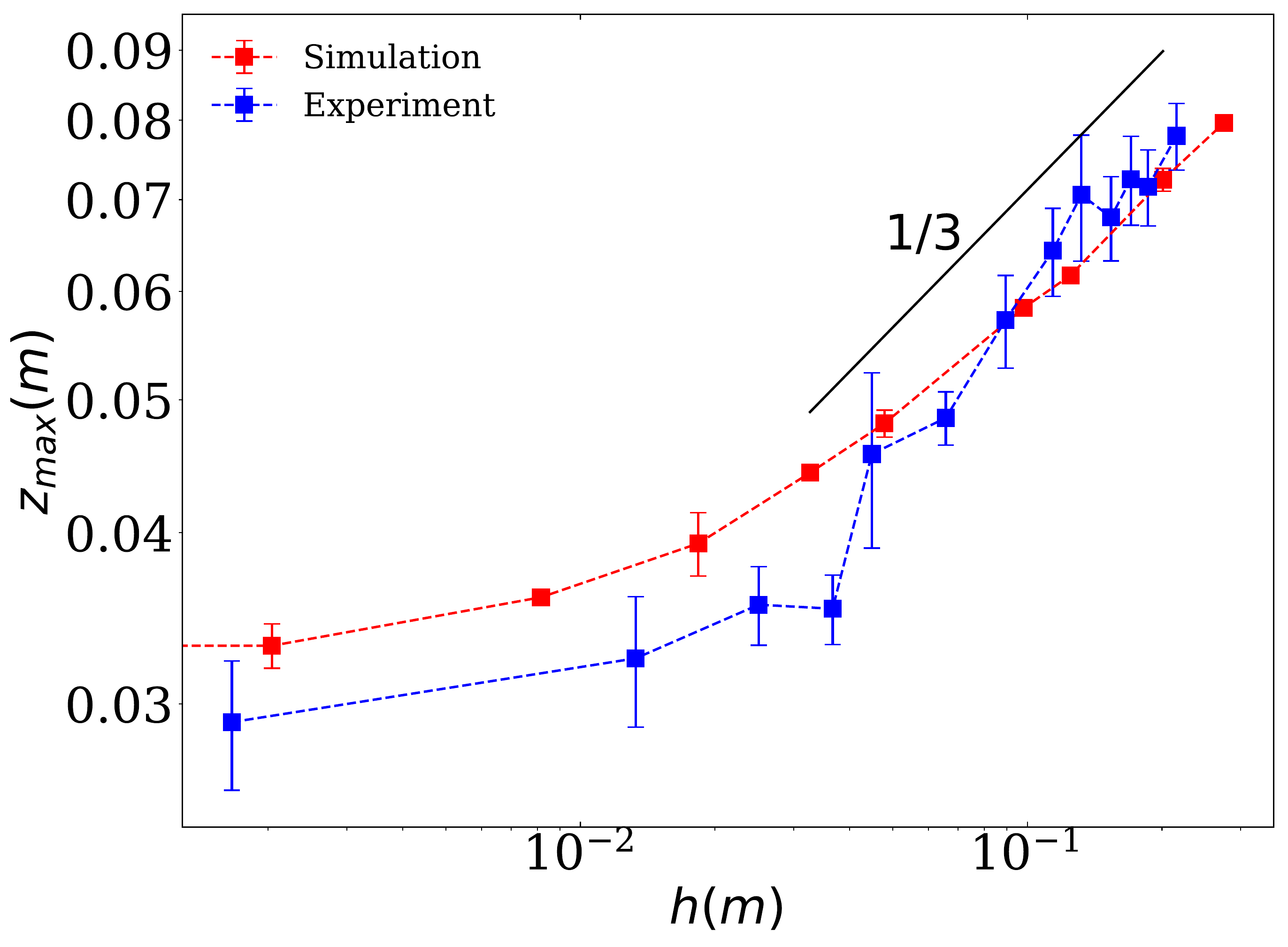} \\
{\large (c)} \\
\includegraphics[width = 0.45\textwidth]{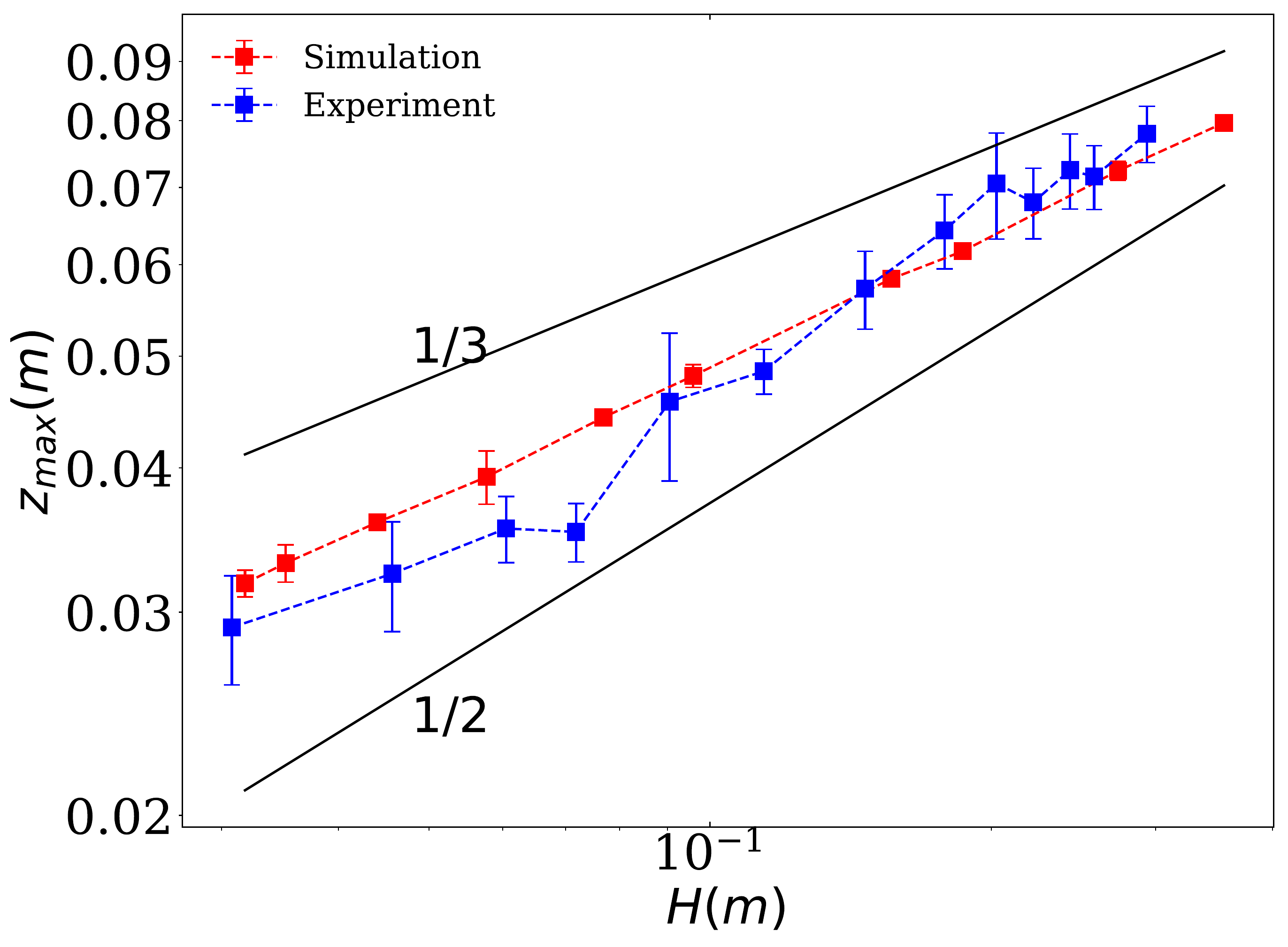} 
\caption{Penetration depth as a function of the impact velocity (a); penetration 
depth as a function of the initial height respect to the surface $h$ in log-log (b); 
penetration depth as a function of the initial height respect to the final position
$H = h+z_{max}$ in log-log (c). In all cases, experimental and numerical outcomes 
are illustrated for comparison.} 
\label{depth}
\end{figure}

In figure \ref{depth}, we focus on the relation between the final penetration depth $z_{max}$ and $v_{\rm i}(0)$. As expected, the final depth increases with increasing initial speed (see Fig.\ref{depth}a). In both experiment and simulations, we find a slightly non-linear response, which is more noticeable for low initial speed. It is important to mention that a linear dependency would be congruent with $z_{max}\propto h^{1/2}$, which was earlier reported and justified as a liquid-like viscous damping \cite{Bruyn2004}. On the other hand, the scaling $z_{max}\propto h^{1/3}$ was also reported in the past \cite{Katsuragi2007,Katsuragi2013} and explained as a complex inertial response of granular beds to impact. For a better comparison, in Figs. \ref{depth}b  and \ref{depth}c, we examine in log-log scale $z_{max}$ vs $h$ and $z_{max}$ vs $H=z_{max}+h$, respectively. As noted, our outcome does not match accurately on any of the two options, but both experiments and simulations are closer to the  $z_{max} \propto H^{1/3}$ dependency, at least within the uncertainties. 

\section*{Response of granular bed}

\subsection*{Coarse-grained fields}

Continuum approaches are also useful, addressing the intruder dynamics and the macroscopic response of the granular bed\cite{Kamrin_2016}. Here, a coarse-graining method \cite{Goldhirsch2010,Weinhart2013,Richard2015,Artoni2019} is also used to examine the local macroscopic responses of the granular bed, as the intruder moves through it. From the $DEM$ data, we compute the macroscopic fields: volume fraction $\varphi\left(\vec{r},t\right) = \rho\left(\vec{r},t\right)/\rho_{\rm p}$, macroscopic velocity $\vec{v}(r,z,t)$ and kinetic stress ${\bf \sigma}_k(r,z,t)$. 

\begin{figure}
\includegraphics[width = \columnwidth]{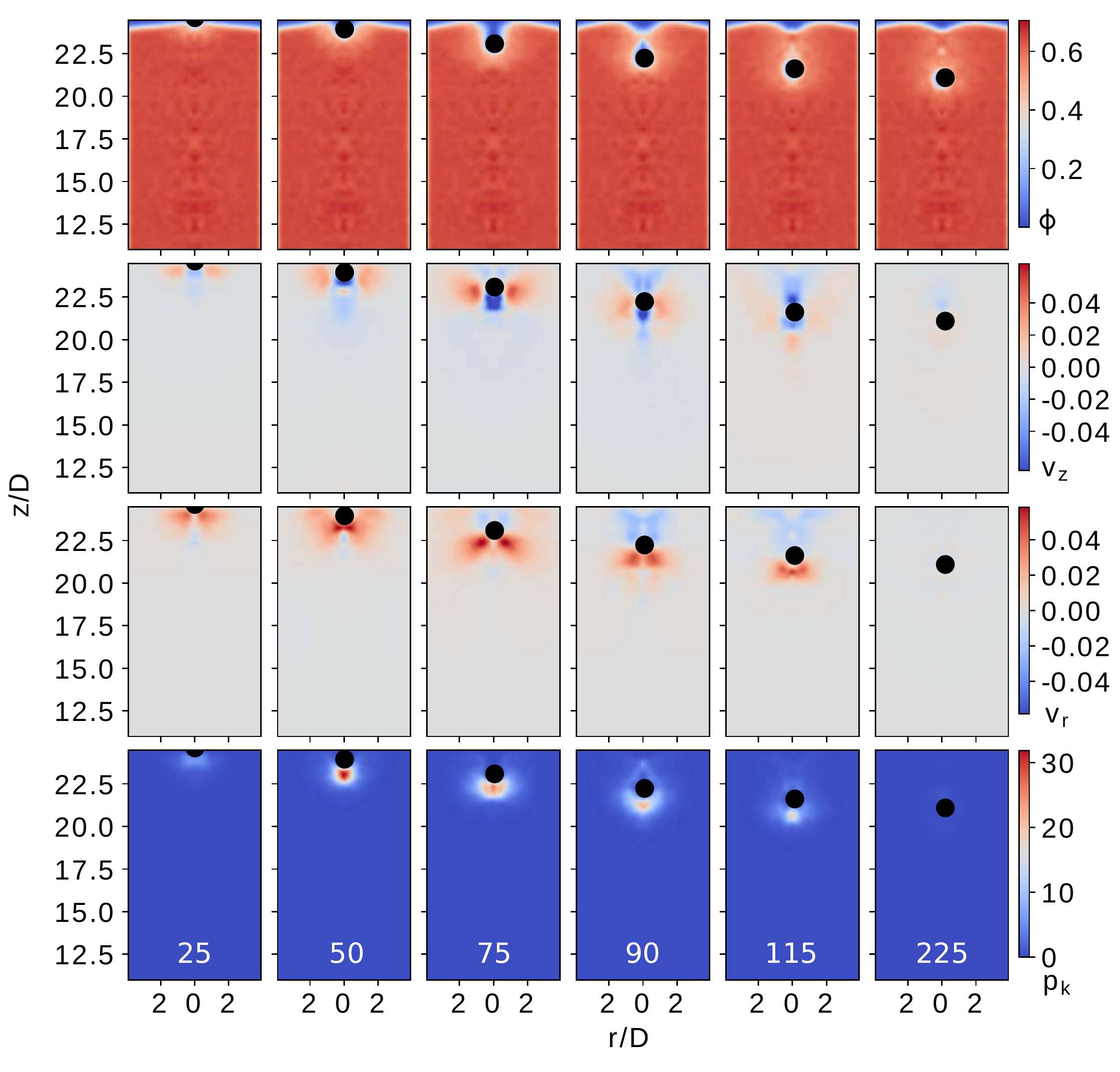} 
\caption{Packing fraction field $\varphi\left(\rho,z,t \right) = \rho\left(\rho,z,t \right)/\rho_{\rm p}$ (row I),
 velocities fields (row II) vertical and (row III) radial; and kinetic pressure fields $p_i$ (row IV), 
 obtained for different times for a initial speed $v_{\rm i}(0) = 0~m/s$. In computation, 
 we use a truncated Gaussian coarse-graining function $\phi(\vec{\mathcal{R}})$ with 
 coarse-grained scale $w = r_p$.} 
\label{slow_fields}
\end{figure}

\begin{figure}
\includegraphics[width = \columnwidth]{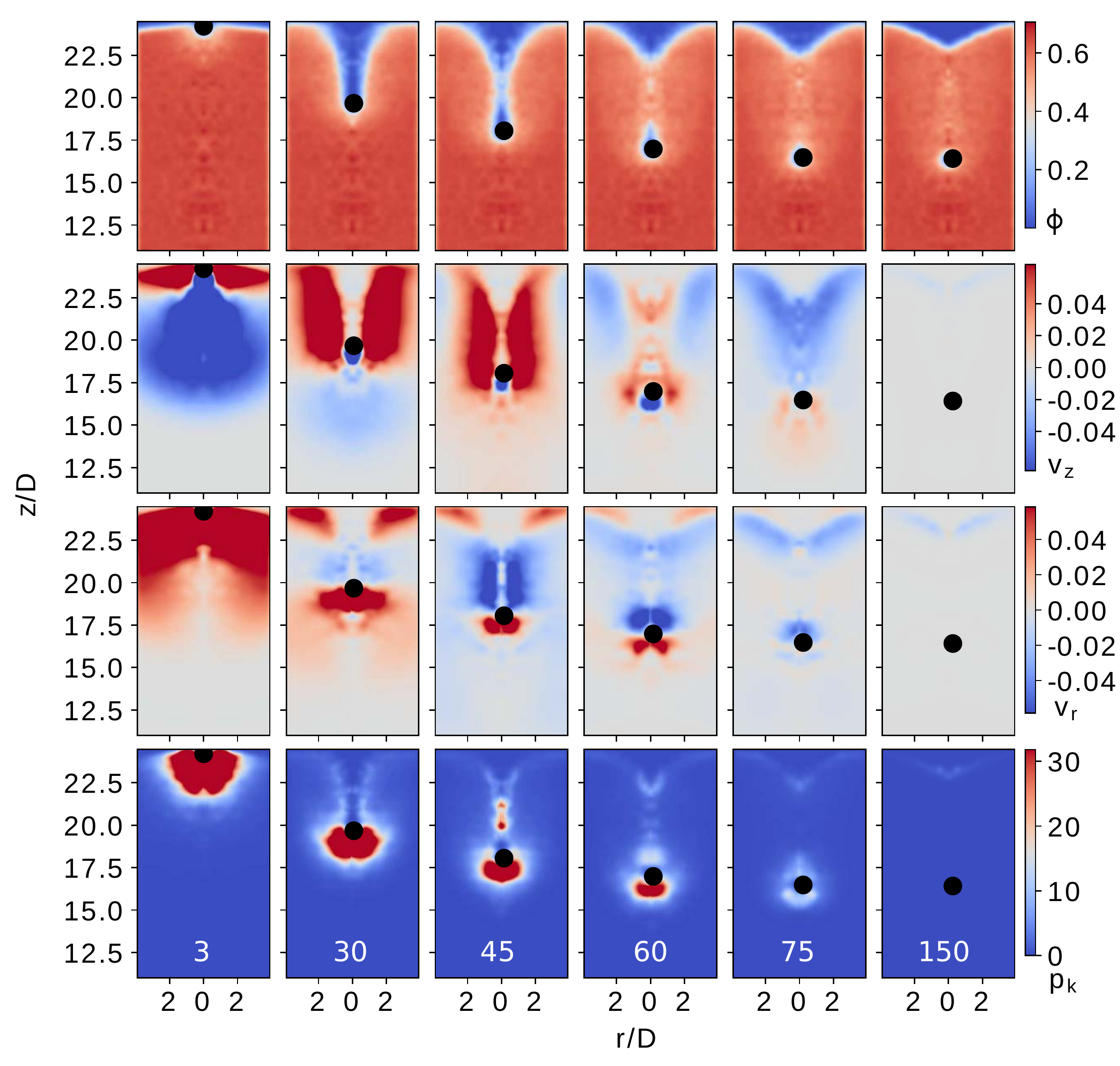} 
\caption{
Packing fraction field $\varphi\left(\vec{r},t \right) = \rho\left(\vec{r},t \right)/\rho_{\rm p}$ (row I), 
velocities fields (row II) vertical and (row III) radial; and kinetic pressure fields $p_k$ (row IV), 
obtained for different times for initial a speed $v_{\rm i}(0) = 2.3~m/s$. The coarse-graining function $\phi(\vec{\mathcal{R}})$ with $w =r_p$ and color-map are the same as in Fig.\ref{slow_fields} for a better comparison.} 
\label{faster_fields}
\end{figure}

In our analysis, we took advantages of the cylindrical symmetry of the system. So that, cylindrical coordinates are optimal to describe it, because the intruder gets into the media by the axis of the cylindrical container, approximately. Thus, we average the vertical and radial components for the studied magnitudes, within an azimuthal representative volume element with the same size.
In Fig.\ref{slow_fields} and Fig.\ref{faster_fields}, outcomes are illustrated for two initial speeds of the intruder, $v_{\rm i}(0) = 0$~m/s and $v_{\rm i}(0) = 2.3$~m/s, respectively.

The coarse-graining methodology \cite{Goldhirsch2010,Weinhart2013,Richard2015,Artoni2019} allows us to reproduce the macroscopic volume fraction fields $\varphi\left(\rho,z,t \right)$ at different times [see color maps in Fig.\,\ref{slow_fields} (row I) and Fig.\,\ref{faster_fields} (row I)] for two extreme initial speeds $v_{\rm i}(0) = 0$  and $v_{\rm i}(0) = 2.32 m/s$, respectively. Note,  
First, it is noticeable that the penetration process does not perturb significantly the macroscopic density, as the impact is more noticeable in the neighborhood of the intruder. Moreover, the fields allow visualizing the time evolution of the void cavity at different depths, as the intruder moves through the system. As noted, the impacting sphere creates a well-defined cavity, which first increases its size and suddenly collapses at a given time $t_{\rm c}$. Remarkably, the cavity shapes obtained in both cases are comparable with previous experiments using X-ray radio-graph \cite{Royer2005,Royer2011,Homan2015}.

The second row in Figs.\,\ref{slow_fields} and \ref{faster_fields} illustrates the vertical velocity field $v_z(\rho,z)$ during the penetration process, while the third row in both figures represents the radial velocity field $v_\rho(\rho,z)$. In general, the intruder notably perturbs the velocity fields of the granular bed, transmitting its mechanical energy through contacting particles over a certain distance. Interestingly, the field $v_z(\rho,z)$ indicates that the existence of a shear band, which correlates with the development of the void cavity. While the intruder penetrates, it pushes away the particles below it towards the bottom and the radial direction of the container. When the intruder travels deep enough, the particles move back to fill the cavity. 
Interesting, the color-maps of $v_z(\rho,z)$ (Figs.\,\ref{slow_fields} (row II) and Figs.\,\ref{faster_fields}  (row II)) also show that, even from the beginning of the process, the intruder notably perturbs the region that is later accessed. Complementary, we quantify the percentage of mobilized contacts (contacts with $F_t>\mu F_n$) in the bottom region of the intruder, after penetrating a distance of $D/2$, and obtaining approximately $4\%$ for the case $v_{\rm i}(0) = 0 $ and $15\%$ for $v_{\rm i}(0) = 2.3~m/s$. These finding clearly indicates that the initial impact perturbs the system at large distances, enhancing its local plasticity and favoring further reordering events. As a result, the granular bed changes its macroscopic response, when varying the initial speed. The latter might explain that the coefficient $\gamma$ (see Eq.(\ref{eq:motion}) intruder's dynamics) is not a material constant, despite the weak changes in macroscopic volume fraction.

Additionally, we also connect the particle local activity to the intruder deceleration, exploring the behavior of the kinetic stress  ${\bf \sigma}^k(\rho,z)$, which is the stress associated with velocity fluctuations (see the fourth row in Figs.\,\ref{slow_fields} and \ref{faster_fields}). Note that momentum transfer as well as energy loss from inelastic collisions is closely associated with the deceleration trajectory of the intruder. This process is very heterogeneous and happens at a length scale comparable to the size of the intruder. The energy consumption rate due particle-particle collisions correlates with the strength of kinetic pressure $p_k(\rho,z)=\rm Tr({\bf \sigma}^k(\rho,z))$. During the impact, it is noticeable the complex propagation of the local kinetic activity, which is quantified by values of the kinetic pressure. As expected, the magnitude of the local activity correlates with the magnitude of the impact velocity in the initial stage of penetration. For the cases with $v_{\rm i}(0)\ne0$, the impact immediately generates a fluidized region bellow the intruder where $p_k(\rho,z)$ maximizes at a certain distance from the intruder. The kinetic stress decays as the intruder decelerates, and other local maximums arise. Interestingly, the time evolution of $p_k(\rho,z)$ indicates that the strength and location of a maximum are invariant, in the reference system of the intruder. For the case of $v_{\rm i}(0)=0$, the fluidized region slowly develops as the intruder penetrates into the system. As the process happens, the general picture is very similar, even though the values of $p_k(\rho,z)$ are one order of magnitude lower. In both cases, the disturbed region ahead of the intruder is on the order of $D$.

\subsection*{Cavity collapse}
\begin{figure}
{\large (a)} \\
\includegraphics[width=0.6\columnwidth]{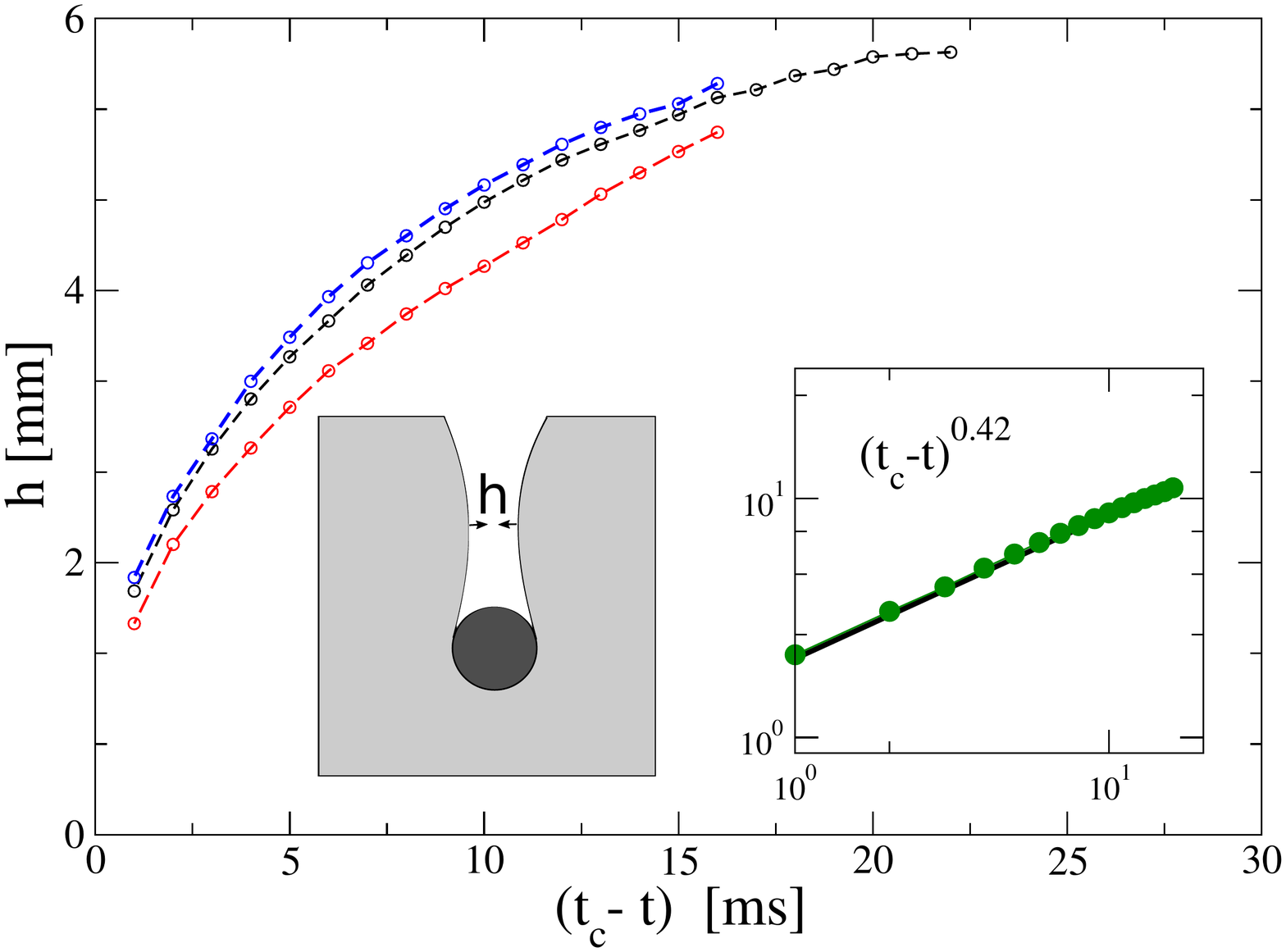} \\
{\large (b)} \\
\includegraphics[width=0.6\columnwidth]{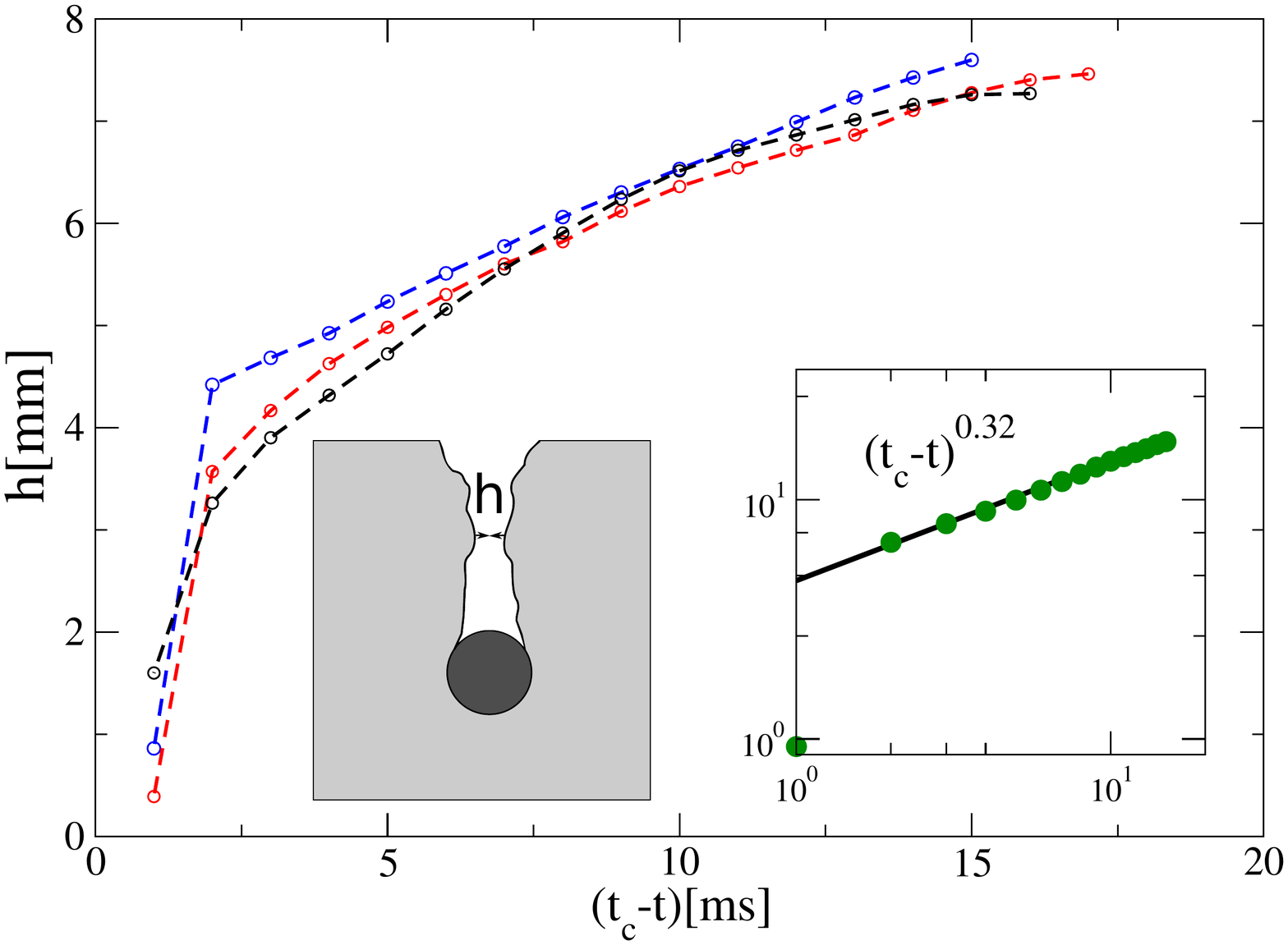}
\caption{Evolution of the cavity radius $h(t)$ towards collapsing, which occurs at $t_{\rm c}$ for the case of $v_{\rm i}(0)=0$ (a) and $v_{\rm i}(0)=2.3$~m/s (b), respectively. Different curves correspond to different realizations of intruder impacts at the same speed and initial packing density. In each case, the insets show the trend of the averaged data of $h(t)$.}
\label{cavity_colapse}
\end{figure}

The computed volume fraction fields allow us to explore nature of void collapse through a quantitative analysis on its dynamics. With that aim, we compute a sequence of $\phi(\rho,z)$ fields every millisecond. Subsequently, void profile $R(z,t)$ is estimated using the definition $\phi(\rho,z)>\phi_{\rm RCP}/2$ for granular bulk.  This definition is plausible given the value of coarse-graining scale $w=r_p$. Following this procedure, the void profile $R(z,t)$ is extracted as a function of time, based on which the dynamics of the cavity radius $h(t)=R(z_{\rm c},t)$, the location $z_{\rm c}$ and time $t_{\rm c}$ of collapsing are identified.  

In the past, the dynamics of the cavity collapse was experimentally accessed using X-Ray Tomography \cite{Royer2005,Homan2015}. It was found that the evolution of the cavity radius $h(t)$ was consistent with a power law $h(t)=(t_{\rm c} - t)^\alpha$ , where $t_{\rm c}$ is the collapsing time, and an exponent $\alpha=0.66$ was reported \cite{Homan2015}. Moreover, several groups have studied theoretically and experimentally the collapsing of air bubbles in liquids \cite{Gordillo2005,Eggers2007,Bergmann2006}. Furthermore, in void collapse due hydrostatic pressure, there are theoretical evidences supporting 
that the cavity radius should diminish as $h(t)=(t_{\rm c} - t)^{1/2}$, resembling a Rayleigh-type singularity \cite{Bergmann2006,Gekle2009}. 

In Fig.\,\ref{cavity_colapse}, the evolution of cavity radius $h(t)$ is illustrated for two different impactor speeds. In general, we find that $h(t)$ smoothly diminishes in the neighborhood of $t_{\rm c}$, exhibiting a power law behavior $h(t)=(t_{\rm c}-t)^{\alpha}$ (see the insets in Fig.\,\ref{cavity_colapse}a and Fig.\,\ref{cavity_colapse}b). This behavior is very similar to the cavity collapse in the wake of an impacting disk on a fluid \cite{Gekle2009}. Moreover, this power law controlled pitch-off was also encountered when a ball is dropped in sand \cite{Homan2015}. We find that the exponent characterizing the collapsing process notably differs depending on the initial speed impact. For the case, $v_{\rm i}(0)=0$ a $\alpha=0.42$ is estimated, while a noticeable lower value $\alpha=0.32$ is found for $v_{\rm i}(0)=2.3$~m/s. Thus, collapsing process gets slower as the initial speed gets larger, which might correlates with the amount of momentum transferred, the size of the perturbed zone and its corresponding dissipation time. It is important to remark, that in the case $v_{\rm i}(0)=2.3$~m/s, the last accessible data point always deviates from this trend, because it is the result of the movement of individual particles.

\subsection*{Characteristic length and time scales}

\begin{figure}
{\large (a)} \\
\includegraphics[width = 0.6\columnwidth]{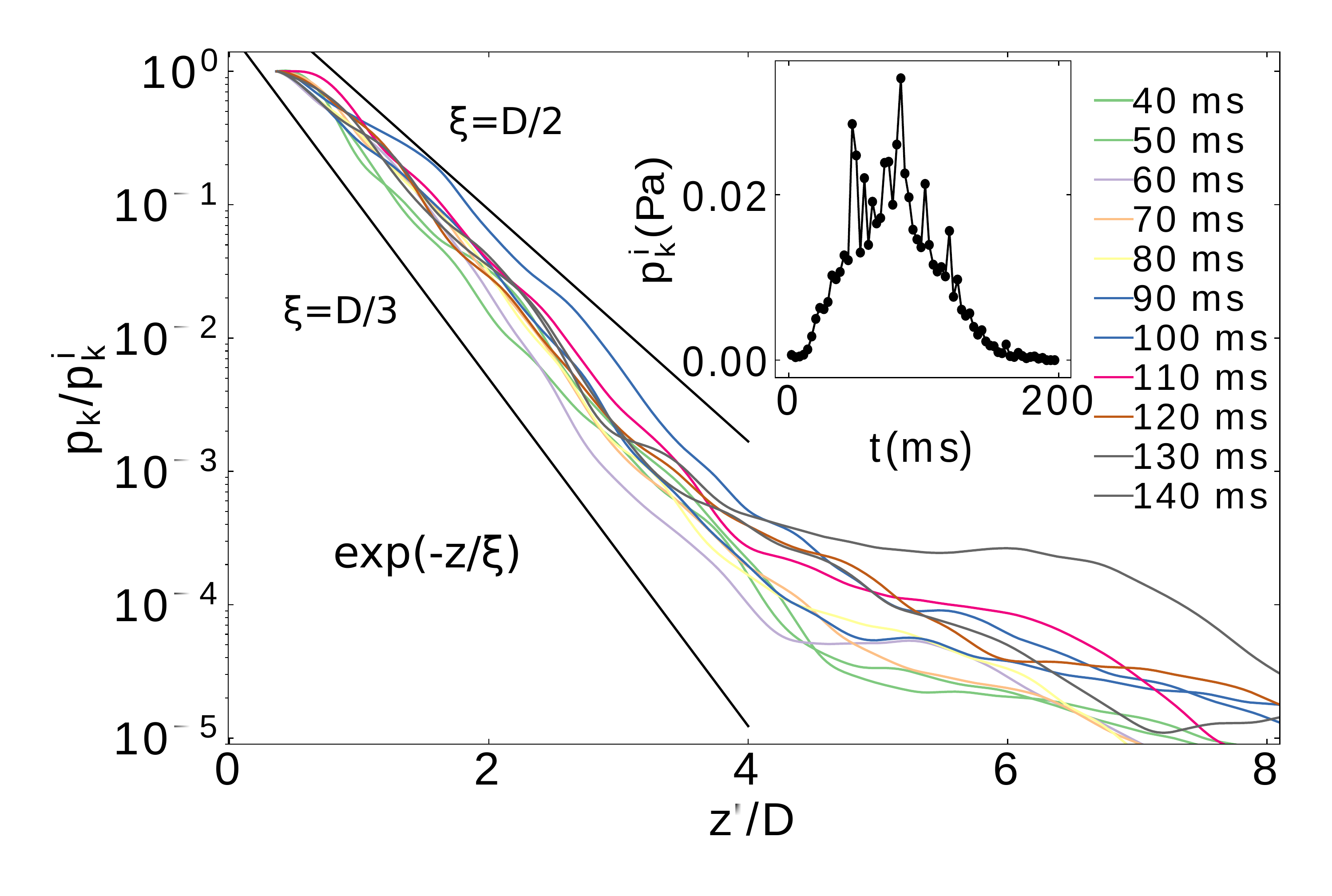} \\
{\large (b)} \\
\includegraphics[width = 0.6\columnwidth]{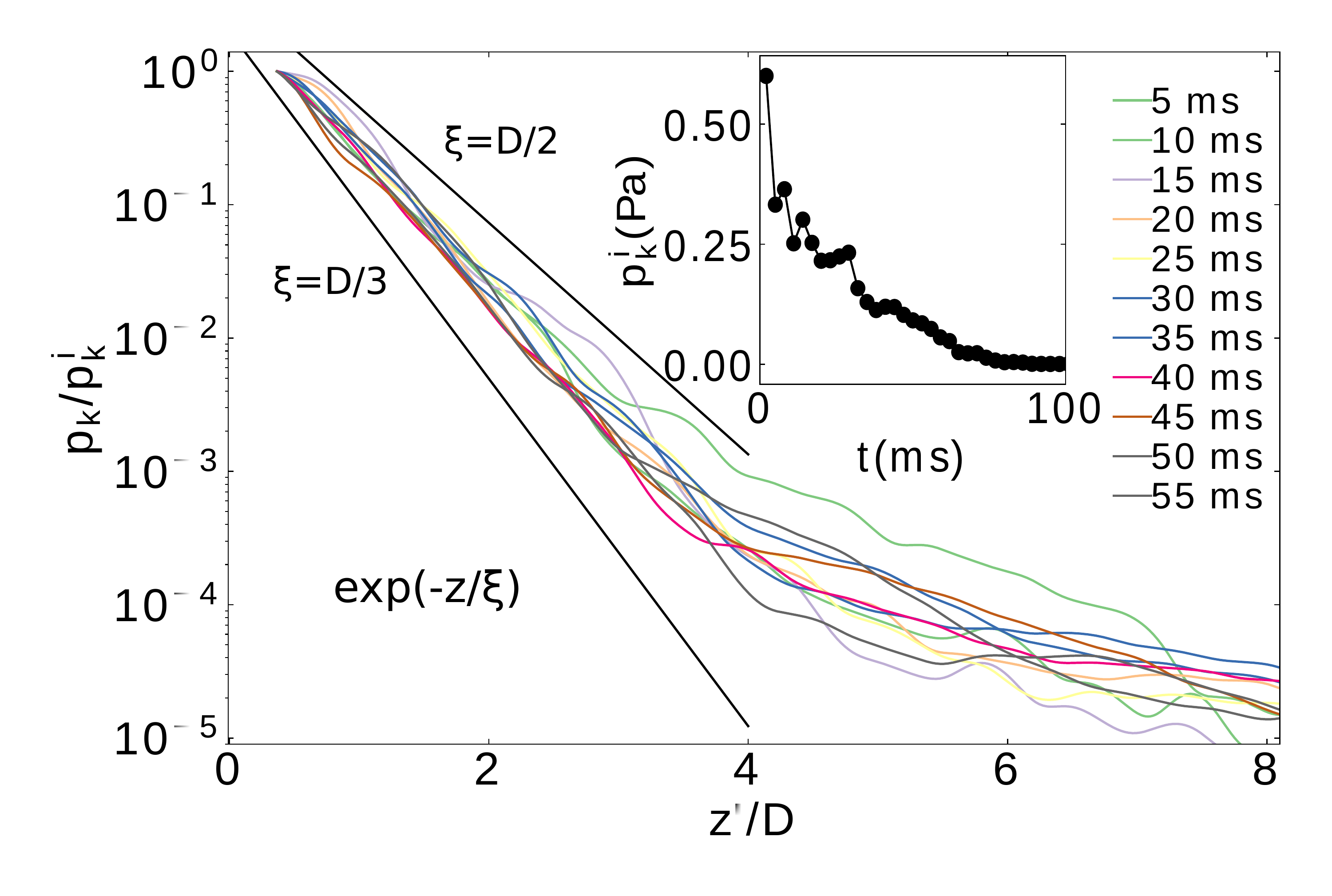} 
\caption{Kinetic pressure profiles $p_k(z')$, obtained in the moving system of the intruder $z'=z-v_{\rm i} t$. a) shows the case $v_{\rm i}(0)=0$ and b) the case $v_{\rm i}(0)=2.3$~m/s. 
In all cases, the data values are re-scaled with their maximum values $p^i_k$ (see insets), which is located in the neighborhood 
of the intruder.}
\label{stress_profiles}
\end{figure}

In order to quantify the characteristic length and time scales of collisional energy dissipation, we examine the kinetic pressure profile in the vertical direction. The averaged values $\langle p_{\rm k}(z) \rangle$ correspond to an average on a radial sector of $\delta r = D$, centered at the intruder location. Fig \ref{stress_profiles} illustrates several results $\langle p_{\rm k}(z) \rangle$ corresponding to two initial intruder speeds and different times;
Fig.\,\ref{stress_profiles}a and Fig.\,\ref{stress_profiles}b, $v_{\rm i}(0)=0$ and $v_{\rm i}(0)=2.3$~m/s, respectively. For a better comparison, the spatial coordinates correspond to the mobile reference system at the intruder $z'=(z-v_{\rm i} t)$ and the values of kinetic pressure are re-scaled with their maximum values $p^i_k$ (see insets of Fig.\,\ref{stress_profiles}a and Fig.\,\ref{stress_profiles}b), which is located in the neighborhood of the intruder. In general, the kinetic stress decays exponentially with the distance respect to the intruder location, regardless of the initial intruder velocity and time. Astonishingly, though the values of $\langle p_{\rm k}(z) \rangle$
 differ in one order of magnitude, the characteristic length scale is practically the same in all cases and it is of order of $D$. In the past, a similar analysis was done for a 2D case\cite{Clark2016}; examining a scaled steady-state velocity field, which was found to decay exponentially, with a correlation length in the order of the particle size.

During the penetration process of an intruder in a granular bed with macroscopic mass density $\rho$, dissipation occurs due to inelastic collisions between grains. In order to build a continuous theoretical approach of this process \cite{Seguin2011}, the starting point is the energy balance equation in terms of the granular temperature $T$, which written in the moving reference system of the intruder $z'=z-v_z(t)t$ reads,    
\begin{equation}
\rho \frac{\partial T}{\partial t} - \rho v_{\rm i} \nabla T = \sigma:{\dot \epsilon}-\nabla \cdot q - \alpha_T T
\label{energy_balance1}
\end{equation}
\noindent 
In equation \ref{energy_balance1}, one can use the Fourier’s law $q = -\lambda_T \nabla T$ for the heat flux with diffusivity $\lambda_T$; while $\alpha_T T$ represents the collisional granular dissipation and ${\dot \epsilon}$ is the macroscopic shear rate. Moreover, $\rho v_{\rm i} \nabla T$ accounts for the convective heat transmission. 

Interestingly, the data shown in Fig.\,\ref{stress_profiles}a and Fig.\ref{stress_profiles}b suggest the presence of a very slow diffusive-convective heat transmission scenario. In that conditions ${\dot \epsilon} \approx 0$ and equation (\ref{energy_balance1}) 
in $1D$ reads as,
\begin{equation}
-\lambda_T \frac{\partial^2 T}{\partial z'^2} + \rho v_{\rm i} \frac{\partial T}{\partial z'} = \alpha_T T + \rho \frac{\partial T}{\partial t}
\label{energy_balance2}
\end{equation}
\noindent
Note that $T(z',t) \cong T_{\rm o} e^{-\frac{t}{\tau}} e^{-\frac{z'}{\xi}}$ is a solution of equation (\ref{energy_balance2}), where $T_{\rm o}$ is the granular temperature at the surface of the intruder. Remark that $\tau$ and $\xi$ are the time and space characteristic values of the penetration process. Naively, $\tau$ is in the order of the penetration time, and  
$\xi$, is the positive root of the characteristic equation $\lambda_T \xi^2-\rho v_{\rm i} \xi + (\alpha_T + \rho/\tau)$. 
From our data and assuming $T(z') \cong p^k(z')$, we find that the length scale of the energy dissipation 
$\xi$ is practically constant $\xi\approx\frac{2}{5}D$. Interesting, this finding is in perfect agreement with the long-scale inertial dissipation type $F_d= -\gamma v_{\rm i}^2$ that we find in all cases. It is important to remark, however, that we find that the dissipation parameter $\gamma$ decreases with increasing initial speed of the intruder. This fact correlates with the initial kinetic pressure that the intruder induces at impact, when mechanical energy is quickly transmitted through the granular bed by means of, e.g., shock waves \cite{Bougie2002,Huang2006}. Thus, right from the beginning the intruder perturbs the strong particle contacts on the force network, and enhances the plasticity of the system. Consequently, the system responses distinctively depending on the initial speed of the intruder. Finally, we note that the existence of a critical length scale ahead of a projectile was also observed in previous investigations on projectile impact on dust agglomerates for understanding the evolution of protoplanetesimal \cite{Guttler2009,Katsuragi_2017}. The authors also showed, based on dimensional analysis, that the characteristic length scale in the system should be directly related to the projectile size. Stepping further, our results show the length scale remains the same, when changing the projectile initial speed and; {\it i.e.} the amount of energy transmitted by projectile to the surrounding particles.

%%%%%%%%%%%%%%%%%%%%%%%%%%%%%%%%%%%%%%%%%%%%%%%%%%%%%%%%%%%%
%                  Conclusions and Outlook
%%%%%%%%%%%%%%%%%%%%%%%%%%%%%%%%%%%%%%%%%%%%%%%%%%%%%%%%%%%%

\section*{Conclusions and outlook}
To summarize, we study experimentally and numerically the penetration of a spherical intruder in a granular bed at various impact velocities using radar particle tracking in experiments and coarse-graining techniques in simulations. The dynamics of the intruder is numerically reproduced and the obtained trajectories are in very good agreement with a well-accepted phenomenological model. However, the obtained coefficient of the inertial drag term, which is $\propto v_{\rm i}^2$, is not a material property, but deviates from a constant value as the initial speed decreases to zero. The latter raises the question of what determines the nature of granular drag. To explain this, we compute the relevant macroscopic fields of the granular bed (volume fraction, macroscopic velocity and the kinetic stress tensor) by means of coarse-graining techniques. Our results show that the intruder perturbs the granular bed even at very long distances, although notable changes in volume fraction are not detected. Further analysis of the cavity collapse behind the intruder reveals a power law diminishes of the neck radius $h(t)$, with an exponent depending on the initial impact speed. 
Moreover, the kinetic pressure profiles $p^k(z')$ in the co-moving frame of the intruder decays always exponentially with the distance to the intruder $z'$, and the characteristic length is practically independent of the initial speed of the intruder. Analytic arguments show that this result indicates a very slow diffusive-convective collisional-energy transmission within the granular bed. In the future, further investigations on how gravity influences granular drag shall be carried out by means of oblique impacts as well as microgravity experiments.

\section*{Method}
\label{sec:meth}

\subsection*{Experimental set-up}

As illustrated in Fig.\,\ref{fig:setup}, the transmission horn antenna emits a $10$\,GHz electromagnetic (EM) wave into the free space, while a spherical projectile moves following an unknown trajectory.  As the dielectric constant of styrofoam particles is very close to air, the granular bed is practically transparent to EM waves. Thus, the mobility of the target with respect to each receiving (Rx) antenna is obtained from the phase shift between the received signal and the emitted one. Finally, a coordinate transformation allows reconstructing the trajectory of the projectile  accurately. Moreover, given the analog nature of the radar system output, an extremely high temporal resolution is also achieved.  The limitation of the temporal signal only arises from the analog-digital (AD) converter. For the normal penetration processes studied in the current investigation,  we use an AD converter (NI-DAQ6015) up to $200$~kHz. Furthermore, the non-invasive nature and the high temporal resolution, make this technique suitable for monitoring the trajectory of spherical and non-spherical projectiles in three dimensions (3D). More details on the radar system as well as its advantages and disadvantages in comparison to other particle imaging techniques can be found in \cite{Ott2017,Huang2019} and \cite{Amon2017}, respectively. 

In the experimental set-up, a steel ball with a radius of $R=0.5$~cm is held by a styrofoam holder at various initial falling height $H$. By gently pulling a thin thread that wrapped around the sphere, we allow the sphere to fall freely with minimized horizontal and angular velocities. The granular sample is composed of EPP particles (Neopolen, P9255) with a slightly ellipsoidal shape and a mean radius of $r_{\rm p}=1.5$~mm, filled into a cylindrical container made of Styrofoam, which is also transparent to EM waves. The radius of the container is $R_{\rm c} = 15$~cm and the filling height is $23$~cm, sufficiently deep to avoid the influence from the side wall. The radar system starts shortly before the projectile release until its final resting state. It operates with a power of $1$~W with the capability of tracking the projectile at a distance of $\sim 1.5$~m with a field of view of $\sim 30$~cm.

\subsection*{DEM and Coarse-Graining Methodology}

We use discrete element modeling (DEM)\cite{poschel05a}, describing a spherical intruder sinking into a granular bed composed of spherical particles. The implementation is a hybrid $CPU/GPU$ algorithm that allows us to efficiently evaluate the dynamics of several hundred thousand particles \cite{cuda10c,Rubio-Largo2016}. We initiate each simulation by generating a random packing of polydisperse spheres with radius $r_{\rm p} = 1.5$~mm, confined in a cylindrical container of radius $R_{\rm c}$ with packing fraction $\phi=0.63\pm0.02$. The spherical intruder is released from the free granular surface with a given initial velocity, the range of which matches that in the experiments.

The $DEM$ algorithm computes the movement of each particle $i=1...N$ for the three translational degrees of freedom, and the rotational movement is described by a quaternion formalism. The interaction force between particle $i$ and particle $j$, is composed of normal and tangential components 
\begin{equation}
{\vec F}_{ij} = {\vec F}_{ij}^{n} + {\vec F}_{ij}^{t}.
\end{equation}
Given the particles normal overlap $\delta_n$ and the tangential relative displacement ${\vec \xi}$, it reads as,
\begin{equation}
{\vec F}_{ij} = (k_n \delta_n + \gamma_n v_r^{n})\times{\hat n} + (k_t \xi + \gamma_t v_r^{t})\times{\hat t},
\end{equation}
Here we used a Hertz–Mindlin model \cite{poschel05a}, $k_n = \frac{4}{3}Y \sqrt{R_e \delta_n}$ and $k_t= 8 G\sqrt{R_e \delta_n}$, $R_e$ is the equivalent radius and $Y$ and $G$ are the equivalent Young’s and shear modulus, respectively. Moreover, $\gamma_n= 2\sqrt{\frac{5}{6}} \beta \sqrt{S_n m^*} $ and $\gamma_t=2\sqrt{\frac{5}{6}} \beta \sqrt{S_t m^*}$, where $S_n= 2Y\sqrt{R \delta_n}$, $S_t=8G\sqrt{R \delta_n}$, 
$m^*=\frac{m_i+m_j}{m_i m_j}$ and $\beta=\frac{\rm ln(e_n)}{\sqrt{\rm ln^2(e_n) + \pi^2}}$. $e_n$ is the normal restitution coefficient of the particles. The tangential relative displacement ${\vec \xi}$ is kept orthogonal to the normal vector and it is truncated as necessary to satisfy the Coulomb constraint $|{\vec F}_{ij}^{t}|<\mu |{\vec F}_{ij}^{n}|$, where $\mu$ is the friction coefficient.  

In all the simulations presented here, the system is composed of $N=180864$ particles, the used contact parameters correspond approximately to particles with a Young's modulus $Y = 800$~kPa~~($G=Y/10$), density $\rho_{\rm p}=92.0~{\rm kg}/{\rm m}^3$ and normal restitution coefficient $e_{\rm n}=0.4$. The intruder radius is $R=0.005~m$ and density $\rho=7800~{\rm kg}/{\rm m}^3$. The particle-wall interaction is modeled using the the same collision parameters used for particle–particle interaction. The parameters are chosen to match the experimental conditions. 

Besides the time evolution of the intruder,  the simulation data also provides the position and velocities of all the particles of the bed, during the entire impact process. Thus, in order to explore the macroscopic response of the system a coarse-graining methodology is also used  \cite{Goldhirsch2010,Weinhart2013,Richard2015}. As a result, the local response of the granular bed, as the intruder moves through it is addressed 
in detail, through  the  macroscopic fields of volume fraction  $\phi(\rho,z)$, particle velocity $\vec{v}(\rho,z)$  and kinetic stress $\sigma_k(\rho,z)$.

According to \cite{Goldhirsch2010,Weinhart2013,Richard2015}, the macroscopic mass density of a granular flow, $\rho(\vec{r})$, at time $t$ is defined by
\begin{equation}
\rho\left(\vec{r},t\right) = \sum_{i=1}^{N} m_i \phi\left(\vec{r}-\vec{r}_i(t)\right)
\label{density_d}
\end{equation}
\noindent where the sum runs over all the particles within the system and $\phi\left(\vec{r}-\vec{r}_i(t)\right)$
is an integrable coarse-graining function. We use a truncated Gaussian coarse-graining function $\phi(\vec{\mathcal{R}}) = A_w e^{-(|\mathcal{R}|/2w)^2}$ 
with cutoff $r_c = 4w$ where the value of $w$ defines the coarse-grained scale. The value of $A_w$ is computed guarantying the normalization condition. From Eq.(\ref{density_d}),
the volume fraction field is $\varphi\left(\vec{r},t \right) = \rho\left(\vec{r},t \right)/\rho_{\rm p}$.

On the other hand, the coarse grained momentum density function $P(\vec{r},t)$ is defined by
\begin{equation}
\vec{P}(\vec{r},t) = \sum_{i=1}^{N} m_i \vec{v}_{i} \phi\left(\vec{r}-\vec{r}_i(t)\right),
\end{equation}
where the $\vec{v}_{i}$ represent the velocity of particle $i$.
The macroscopic velocity field $\vec{V}(\vec{r},t)$ is then defined as the ratio of 
momentum and density fields,
\begin{equation}
\label{velocidad}
\vec{V}(\vec{r},t) = \vec{P}(\vec{r},t)/\rho(\vec{r},t).
\end{equation}

Additionally, the propagation of the kinetic activity in the particle bed is quantified by values of the kinetic stress tensor, 
which reads,
\begin{eqnarray}
\mathbf{\sigma^k} \left(\vec{r},t\right) = \sum_{i=1}^{N} m_i \vec{v}_{i }^{\prime} \otimes \vec{v}_{i }^{\prime} \phi\left(\vec{r}-\vec{r}_i(t)\right),
\label{kinetic_stress}
\end{eqnarray}
\noindent 
where $\vec{v}_i^{\prime}$ is fluctuation of the velocity of particle $i$, respect to the mean field
$\vec{v}_{i}^{\prime} (t,\vec{r}) = \vec{v}_{i} (t) - \vec{V} (\vec{r},t)$.

It is generally accepted that the trace of kinetic stress tensor (kinetic pressure) is proportional to the granular
temperature. Thus, it represents a measurement of the grain fluctuations respect to the mean velocity field.   
That's why it can be used to examine the propagation of the kinetic activity as the intruder moves through it.

\bibliography{radar}

\begin{thebibliography}{10}
\expandafter\ifx\csname url\endcsname\relax
  \def\url#1{\texttt{#1}}\fi
\expandafter\ifx\csname urlprefix\endcsname\relax\def\urlprefix{URL }\fi
\expandafter\ifx\csname doiprefix\endcsname\relax\def\doiprefix{DOI }\fi
\providecommand{\bibinfo}[2]{#2}
\providecommand{\eprint}[2][]{\url{#2}}

\bibitem{Chhabra2006}
\bibinfo{author}{Chhabra, R.~P.}
\newblock \emph{\bibinfo{title}{Bubbles, {Drops}, and {Particles} in
  {Non}-{Newtonian} {Fluids}, {Second} {Edition}}} (\bibinfo{publisher}{CRC
  Press}, \bibinfo{year}{2006}).
\newblock \bibinfo{note}{Google-Books-ID: ekIrBgAAQBAJ}.

\bibitem{Katsuragi2016}
\bibinfo{author}{Katsuragi, H.}
\newblock \emph{\bibinfo{title}{Physics of {Soft} {Impact} and {Cratering}}},
  vol. \bibinfo{volume}{910} of \emph{\bibinfo{series}{Lecture {Notes} in
  {Physics}}} (\bibinfo{publisher}{Springer Japan}, \bibinfo{address}{Tokyo},
  \bibinfo{year}{2016}).

\bibitem{Meer2017}
\bibinfo{author}{van~der Meer, D.}
\newblock \bibinfo{title}{Impact on {Granular} {Beds}}.
\newblock \emph{\bibinfo{journal}{Annual Review of Fluid Mechanics}}
  \textbf{\bibinfo{volume}{49}}, \bibinfo{pages}{463--484}
  (\bibinfo{year}{2017}).

\bibitem{Ruiz2013}
\bibinfo{author}{Ruiz-Suarez, J.~C.}
\newblock \bibinfo{title}{Penetration of projectiles into granular targets}.
\newblock \emph{\bibinfo{journal}{Reports on Progress in Physics}}
  \textbf{\bibinfo{volume}{76}}, \bibinfo{pages}{066601}
  (\bibinfo{year}{2013}).

\bibitem{Hosoi2015}
\bibinfo{author}{Hosoi, A.} \& \bibinfo{author}{Goldman, D.~I.}
\newblock \bibinfo{title}{Beneath {Our} {Feet}: {Strategies} for {Locomotion}
  in {Granular} {Media}}.
\newblock \emph{\bibinfo{journal}{Annual Review of Fluid Mechanics}}
  \textbf{\bibinfo{volume}{47}}, \bibinfo{pages}{431--453}
  (\bibinfo{year}{2015}).

\bibitem{Robins1742}
\bibinfo{author}{Robins, B.}
\newblock \emph{\bibinfo{title}{New {Principles} of {Gunnery}}}
  (\bibinfo{address}{London}, \bibinfo{year}{1742}).

\bibitem{Poncelet1829}
\bibinfo{author}{Poncelet, J.~V.}
\newblock \emph{\bibinfo{title}{Cours de M\'ecanique Industrielle}}
  (\bibinfo{address}{Paris}, \bibinfo{year}{1829}).

\bibitem{Uehara2003}
\bibinfo{author}{Uehara, J.~S.}, \bibinfo{author}{Ambroso, M.~A.},
  \bibinfo{author}{Ojha, R.~P.} \& \bibinfo{author}{Durian, D.~J.}
\newblock \bibinfo{title}{Low-{Speed} {Impact} {Craters} in {Loose} {Granular}
  {Media}}.
\newblock \emph{\bibinfo{journal}{Physical Review Letters}}
  \textbf{\bibinfo{volume}{90}}, \bibinfo{pages}{194301}
  (\bibinfo{year}{2003}).

\bibitem{Bruyn2004}
\bibinfo{author}{de~Bruyn, J.~R.} \& \bibinfo{author}{Walsh, A.~M.}
\newblock \bibinfo{title}{Penetration of spheres into loose granular media}.
\newblock \emph{\bibinfo{journal}{Canadian Journal of Physics}}
  \textbf{\bibinfo{volume}{82}}, \bibinfo{pages}{439--446}
  (\bibinfo{year}{2004}).

\bibitem{Katsuragi2007}
\bibinfo{author}{Katsuragi, H.} \& \bibinfo{author}{Durian, D.~J.}
\newblock \bibinfo{title}{Unified force law for granular impact cratering}.
\newblock \emph{\bibinfo{journal}{Nature Physics}}
  \textbf{\bibinfo{volume}{3}}, \bibinfo{pages}{420--423}
  (\bibinfo{year}{2007}).

\bibitem{Brzinski2013}
\bibinfo{author}{Brzinski~III, T.~A.}, \bibinfo{author}{Mayor, P.} \&
  \bibinfo{author}{Durian, D.~J.}
\newblock \bibinfo{title}{Depth-{Dependent} {Resistance} of {Granular} {Media}
  to {Vertical} {Penetration}}.
\newblock \emph{\bibinfo{journal}{Physical Review Letters}}
  \textbf{\bibinfo{volume}{111}}, \bibinfo{pages}{168002}
  (\bibinfo{year}{2013}).

\bibitem{Colaprete2010}
\bibinfo{author}{Colaprete, A.} \emph{et~al.}
\newblock \bibinfo{title}{Detection of {Water} in the {LCROSS} {Ejecta}
  {Plume}}.
\newblock \emph{\bibinfo{journal}{Science}} \textbf{\bibinfo{volume}{330}},
  \bibinfo{pages}{463--468} (\bibinfo{year}{2010}).

\bibitem{Shinbrot2017}
\bibinfo{author}{Shinbrot, T.}, \bibinfo{author}{Sabuwala, T.},
  \bibinfo{author}{Siu, T.}, \bibinfo{author}{Vivar~Lazo, M.} \&
  \bibinfo{author}{Chakraborty, P.}
\newblock \bibinfo{title}{Size {Sorting} on the {Rubble}-{Pile} {Asteroid}
  {Itokawa}}.
\newblock \emph{\bibinfo{journal}{Physical Review Letters}}
  \textbf{\bibinfo{volume}{118}}, \bibinfo{pages}{111101}
  (\bibinfo{year}{2017}).

\bibitem{Royer2005}
\bibinfo{author}{Royer, J.~R.} \emph{et~al.}
\newblock \bibinfo{title}{Formation of granular jets observed by high-speed
  x-ray radiography}.
\newblock \emph{\bibinfo{journal}{Nature Physics}}
  \textbf{\bibinfo{volume}{1}}, \bibinfo{pages}{164--167}
  (\bibinfo{year}{2005}).

\bibitem{Askari2016}
\bibinfo{author}{Askari, H.} \& \bibinfo{author}{Kamrin, K.}
\newblock \bibinfo{title}{Intrusion rheology in grains and other
  flowableÂ materials}.
\newblock \emph{\bibinfo{journal}{Nature Materials}}
  \textbf{\bibinfo{volume}{25}}, \bibinfo{pages}{1274} (\bibinfo{year}{2016}).

\bibitem{Kang2018}
\bibinfo{author}{Kang, W.}, \bibinfo{author}{Feng, Y.}, \bibinfo{author}{Liu,
  C.} \& \bibinfo{author}{Blumenfeld, R.}
\newblock \bibinfo{title}{Archimedes's law explains penetration of solids into
  granular media}.
\newblock \emph{\bibinfo{journal}{Nature Communications}}
  \textbf{\bibinfo{volume}{9}}, \bibinfo{pages}{1101} (\bibinfo{year}{2018}).

\bibitem{Altshuler2014}
\bibinfo{author}{Altshuler, E.} \emph{et~al.}
\newblock \bibinfo{title}{Settling into dry granular media in different
  gravities}.
\newblock \emph{\bibinfo{journal}{Geophysical Research Letters}}
  \textbf{\bibinfo{volume}{41}}, \bibinfo{pages}{3032--3037}
  (\bibinfo{year}{2014}).

\bibitem{Pacheco2010}
\bibinfo{author}{Pacheco-Vazquez, F.} \& \bibinfo{author}{Ruiz-Suarez, J.}
\newblock \bibinfo{title}{Cooperative dynamics in the penetration of a group of
  intruders in a granular medium}.
\newblock \emph{\bibinfo{journal}{Nature Communications}}
  \textbf{\bibinfo{volume}{1}}, \bibinfo{pages}{123} (\bibinfo{year}{2010}).

\bibitem{Tsimring2005}
\bibinfo{author}{Tsimring, L.} \& \bibinfo{author}{Volfson, D.}
\newblock \bibinfo{title}{Modeling of impact cratering in granular media}.
\newblock \emph{\bibinfo{journal}{Powders and Grins}}
  \textbf{\bibinfo{volume}{2}}, \bibinfo{pages}{1215} (\bibinfo{year}{2005}).

\bibitem{Pacheco2011}
\bibinfo{author}{Pacheco-Vazquez, F.} \emph{et~al.}
\newblock \bibinfo{title}{Infinite {Penetration} of a {Projectile} into a
  {Granular} {Medium}}.
\newblock \emph{\bibinfo{journal}{Physical Review Letters}}
  \textbf{\bibinfo{volume}{106}}, \bibinfo{pages}{218001}
  (\bibinfo{year}{2011}).

\bibitem{Lopez2017}
\bibinfo{author}{L\'{o}pez-Rodr\'{i}guez, L.~A.} \&
  \bibinfo{author}{Pacheco-V\'{a}zquez, F.}
\newblock \bibinfo{title}{Friction force regimes and the conditions for endless
  penetration of an intruder into a granular medium}.
\newblock \emph{\bibinfo{journal}{Physical Review E}}
  \textbf{\bibinfo{volume}{96}}, \bibinfo{pages}{030901}
  (\bibinfo{year}{2017}).

\bibitem{Seguin2017}
\bibinfo{author}{Seguin, A.} \& \bibinfo{author}{Gondret, P.}
\newblock \bibinfo{title}{Drag force in a cold or hot granular medium}.
\newblock \emph{\bibinfo{journal}{Physical Review E}}
  \textbf{\bibinfo{volume}{96}}, \bibinfo{pages}{032905}
  (\bibinfo{year}{2017}).

\bibitem{Crassous2007}
\bibinfo{author}{Crassous, J.}, \bibinfo{author}{Beladjine, D.} \&
  \bibinfo{author}{Valance, A.}
\newblock \bibinfo{title}{Impact of a projectile on a granular medium described
  by a collision model}.
\newblock \emph{\bibinfo{journal}{Phys. Rev. Lett.}}
  \textbf{\bibinfo{volume}{99}}, \bibinfo{pages}{248001}
  (\bibinfo{year}{2007}).

\bibitem{Seguin2009}
\bibinfo{author}{Seguin, A.}, \bibinfo{author}{Bertho, Y.},
  \bibinfo{author}{Gondret, P.} \& \bibinfo{author}{Crassous, J.}
\newblock \bibinfo{title}{Sphere penetration by impact in a granular medium: A
  collisional process}.
\newblock \emph{\bibinfo{journal}{{EPL} (Europhysics Letters)}}
  \textbf{\bibinfo{volume}{88}}, \bibinfo{pages}{44002} (\bibinfo{year}{2009}).

\bibitem{Clark2012}
\bibinfo{author}{Clark, A.~H.}, \bibinfo{author}{Kondic, L.} \&
  \bibinfo{author}{Behringer, R.~P.}
\newblock \bibinfo{title}{Particle {Scale} {Dynamics} in {Granular} {Impact}}.
\newblock \emph{\bibinfo{journal}{Physical Review Letters}}
  \textbf{\bibinfo{volume}{109}}, \bibinfo{pages}{238302}
  (\bibinfo{year}{2012}).

\bibitem{Takehara2014}
\bibinfo{author}{Takehara, Y.} \& \bibinfo{author}{Okumura, K.}
\newblock \bibinfo{title}{High-velocity drag friction in granular media near
  the jamming point}.
\newblock \emph{\bibinfo{journal}{Phys. Rev. Lett.}}
  \textbf{\bibinfo{volume}{112}}, \bibinfo{pages}{148001}
  (\bibinfo{year}{2014}).

\bibitem{Jaeger1996}
\bibinfo{author}{Jaeger, H.~M.}, \bibinfo{author}{Nagel, S.~R.} \&
  \bibinfo{author}{Behringer, R.~P.}
\newblock \bibinfo{title}{Granular solids, liquids, and gases}.
\newblock \emph{\bibinfo{journal}{Rev. Mod. Phys.}}
  \textbf{\bibinfo{volume}{68}}, \bibinfo{pages}{1259} (\bibinfo{year}{1996}).

\bibitem{Duran2000}
\bibinfo{author}{Duran, J.}
\newblock \emph{\bibinfo{title}{Sands, Powders and Grains (An Introduction to
  the Physics of Granular Materials)}} (\bibinfo{publisher}{Springer-Verlag},
  \bibinfo{address}{New York}, \bibinfo{year}{2000}), \bibinfo{edition}{1} edn.

\bibitem{Basf}
\bibinfo{title}{https://products.basf.com/en/neopolen.html}.

\bibitem{Zarm}
\bibinfo{title}{https://www.zarm.uni-bremen.de/en/}.

\bibitem{Ott2017}
\bibinfo{author}{Ott, F.}, \bibinfo{author}{Herminghaus, S.} \&
  \bibinfo{author}{Huang, K.}
\newblock \bibinfo{title}{Radar for tracer particles}.
\newblock \emph{\bibinfo{journal}{Rev. Sci. Instrum.}}
  \textbf{\bibinfo{volume}{88}}, \bibinfo{pages}{051801}
  (\bibinfo{year}{2017}).

\bibitem{Ciamarra2004}
\bibinfo{author}{Pica~Ciamarra, M.} \emph{et~al.}
\newblock \bibinfo{title}{Dynamics of {Drag} and {Force} {Distributions} for
  {Projectile} {Impact} in a {Granular} {Medium}}.
\newblock \emph{\bibinfo{journal}{Physical Review Letters}}
  \textbf{\bibinfo{volume}{92}}, \bibinfo{pages}{194301}
  (\bibinfo{year}{2004}).

\bibitem{Katsuragi2013}
\bibinfo{author}{Katsuragi, H.} \& \bibinfo{author}{Durian, D.~J.}
\newblock \bibinfo{title}{Drag force scaling for penetration into granular
  media}.
\newblock \emph{\bibinfo{journal}{Physical Review E}}
  \textbf{\bibinfo{volume}{87}}, \bibinfo{pages}{052208}
  (\bibinfo{year}{2013}).

\bibitem{Xu2014}
\bibinfo{author}{Xu, Y.}, \bibinfo{author}{Padding, J.~T.} \&
  \bibinfo{author}{Kuipers, J. A.~M.}
\newblock \bibinfo{title}{Numerical investigation of the vertical plunging
  force of a spherical intruder into a prefluidized granular bed}.
\newblock \emph{\bibinfo{journal}{Physical Review E}}
  \textbf{\bibinfo{volume}{90}}, \bibinfo{pages}{062203}
  (\bibinfo{year}{2014}).

\bibitem{Clark_2013}
\bibinfo{author}{Clark, A.~H.} \& \bibinfo{author}{Behringer, R.~P.}
\newblock \bibinfo{title}{Granular impact model as an energy-depth relation}.
\newblock \emph{\bibinfo{journal}{{EPL} (Europhysics Letters)}}
  \textbf{\bibinfo{volume}{101}}, \bibinfo{pages}{64001}
  (\bibinfo{year}{2013}).

\bibitem{Bester2017}
\bibinfo{author}{Bester, C.~S.} \& \bibinfo{author}{Behringer, R.~P.}
\newblock \bibinfo{title}{Collisional model of energy dissipation in
  three-dimensional granular impact}.
\newblock \emph{\bibinfo{journal}{Phys. Rev. E}} \textbf{\bibinfo{volume}{95}},
  \bibinfo{pages}{032906} (\bibinfo{year}{2017}).

\bibitem{Kamrin_2016}
\bibinfo{author}{Dunatunga, S.} \& \bibinfo{author}{Kamrin, K.}
\newblock \bibinfo{title}{Continuum modeling of projectile impact and
  penetration in dry granular media}.
\newblock \emph{\bibinfo{journal}{Journal of the Mechanics and Physics of
  Solids}} \textbf{\bibinfo{volume}{100}}, \bibinfo{pages}{45 -- 60}
  (\bibinfo{year}{2017}).

\bibitem{Goldhirsch2010}
\bibinfo{author}{Goldhirsch, I.}
\newblock \bibinfo{title}{Stress, stress asymmetry and couple stress: from
  discrete particles to continuous fields}.
\newblock \emph{\bibinfo{journal}{Granular Matter}}
  \textbf{\bibinfo{volume}{12}}, \bibinfo{pages}{239--252}
  (\bibinfo{year}{2010}).

\bibitem{Weinhart2013}
\bibinfo{author}{Weinhart, T.}, \bibinfo{author}{Hartkamp, R.},
  \bibinfo{author}{Thornton, A.~R.} \& \bibinfo{author}{Luding, S.}
\newblock \bibinfo{title}{Coarse-grained local and objective continuum
  description of three-dimensional granular flows down an inclined surface}.
\newblock \emph{\bibinfo{journal}{Physics of Fluids}}
  \textbf{\bibinfo{volume}{25}} (\bibinfo{year}{2013}).

\bibitem{Richard2015}
\bibinfo{author}{Artoni, R.} \& \bibinfo{author}{Richard, P.}
\newblock \bibinfo{title}{Average balance equations, scale dependence, and
  energy cascade for granular materials}.
\newblock \emph{\bibinfo{journal}{Phys. Rev. E}} \textbf{\bibinfo{volume}{91}},
  \bibinfo{pages}{032202} (\bibinfo{year}{2015}).

\bibitem{Artoni2019}
\bibinfo{author}{Artoni, R.} \& \bibinfo{author}{Richard, P.}
\newblock \bibinfo{title}{Coarse graining for granular materials: micro-polar
  balances}.
\newblock \emph{\bibinfo{journal}{Acta Mech}} \textbf{\bibinfo{volume}{230}},
  \bibinfo{pages}{3055--3069} (\bibinfo{year}{2019}).

\bibitem{Royer2011}
\bibinfo{author}{Royer, J.~R.}, \bibinfo{author}{Conyers, B.},
  \bibinfo{author}{Corwin, E.~I.}, \bibinfo{author}{Eng, P.~J.} \&
  \bibinfo{author}{Jaeger, H.~M.}
\newblock \bibinfo{title}{The role of interstitial gas in determining the
  impact response of granular beds}.
\newblock \emph{\bibinfo{journal}{EPL (Europhysics Letters)}}
  \textbf{\bibinfo{volume}{93}}, \bibinfo{pages}{28008} (\bibinfo{year}{2011}).

\bibitem{Homan2015}
\bibinfo{author}{Homan, T.}, \bibinfo{author}{Mudde, R.},
  \bibinfo{author}{Lohse, D.} \& \bibinfo{author}{Meer, D. v.~d.}
\newblock \bibinfo{title}{High-speed {X}-ray imaging of a ball impacting on
  loose sand}.
\newblock \emph{\bibinfo{journal}{Journal of Fluid Mechanics}}
  \textbf{\bibinfo{volume}{777}}, \bibinfo{pages}{690--706}
  (\bibinfo{year}{2015}).

\bibitem{Gordillo2005}
\bibinfo{author}{Gordillo, J.~M.}, \bibinfo{author}{Sevilla, A.},
  \bibinfo{author}{Rodr\'{\i}guez-Rodr\'{\i}guez, J.} \&
  \bibinfo{author}{Mart\'{\i}nez-Baz\'an, C.}
\newblock \bibinfo{title}{Axisymmetric bubble pinch-off at high reynolds
  numbers}.
\newblock \emph{\bibinfo{journal}{Phys. Rev. Lett.}}
  \textbf{\bibinfo{volume}{95}}, \bibinfo{pages}{194501}
  (\bibinfo{year}{2005}).

\bibitem{Eggers2007}
\bibinfo{author}{Eggers, J.}, \bibinfo{author}{Fontelos, M.~A.},
  \bibinfo{author}{Leppinen, D.} \& \bibinfo{author}{Snoeijer, J.~H.}
\newblock \bibinfo{title}{Theory of the collapsing axisymmetric cavity}.
\newblock \emph{\bibinfo{journal}{Phys. Rev. Lett.}}
  \textbf{\bibinfo{volume}{98}}, \bibinfo{pages}{094502}
  (\bibinfo{year}{2007}).

\bibitem{Bergmann2006}
\bibinfo{author}{Bergmann, R.} \emph{et~al.}
\newblock \bibinfo{title}{Giant bubble pinch-off}.
\newblock \emph{\bibinfo{journal}{Phys. Rev. Lett.}}
  \textbf{\bibinfo{volume}{96}}, \bibinfo{pages}{154505}
  (\bibinfo{year}{2006}).

\bibitem{Gekle2009}
\bibinfo{author}{Gekle, S.}, \bibinfo{author}{Snoeijer, J.~H.},
  \bibinfo{author}{Lohse, D.} \& \bibinfo{author}{van~der Meer, D.}
\newblock \bibinfo{title}{Approach to universality in axisymmetric bubble
  pinch-off}.
\newblock \emph{\bibinfo{journal}{Phys. Rev. E}} \textbf{\bibinfo{volume}{80}},
  \bibinfo{pages}{036305} (\bibinfo{year}{2009}).

\bibitem{Clark2016}
\bibinfo{author}{Clark, A.~H.}, \bibinfo{author}{Kondic, L.} \&
  \bibinfo{author}{Behringer, R.~P.}
\newblock \bibinfo{title}{Steady flow dynamics during granular impact}.
\newblock \emph{\bibinfo{journal}{Phys. Rev. E}} \textbf{\bibinfo{volume}{93}},
  \bibinfo{pages}{050901} (\bibinfo{year}{2016}).

\bibitem{Seguin2011}
\bibinfo{author}{Seguin, A.}, \bibinfo{author}{Bertho, Y.},
  \bibinfo{author}{Gondret, P.} \& \bibinfo{author}{Crassous, J.}
\newblock \bibinfo{title}{Dense granular flow around a penetrating object:
  Experiment and hydrodynamic model}.
\newblock \emph{\bibinfo{journal}{Phys. Rev. Lett.}}
  \textbf{\bibinfo{volume}{107}}, \bibinfo{pages}{048001}
  (\bibinfo{year}{2011}).

\bibitem{Bougie2002}
\bibinfo{author}{Bougie, J.}, \bibinfo{author}{Moon, S.~J.},
  \bibinfo{author}{Swift, J.~B.} \& \bibinfo{author}{Swinney, H.~L.}
\newblock \bibinfo{title}{Shocks in vertically oscillated granular layers}.
\newblock \emph{\bibinfo{journal}{Phys. Rev. E}} \textbf{\bibinfo{volume}{66}},
  \bibinfo{pages}{051301} (\bibinfo{year}{2002}).

\bibitem{Huang2006}
\bibinfo{author}{Huang, K.}, \bibinfo{author}{Zhang, P.},
  \bibinfo{author}{Miao, G.} \& \bibinfo{author}{Wei, R.}
\newblock \bibinfo{title}{Dynamic behaviors of supersonic granular media under
  vertical vibration}.
\newblock \emph{\bibinfo{journal}{Ultrasonics}} \textbf{\bibinfo{volume}{44}},
  \bibinfo{pages}{e1487--e1489} (\bibinfo{year}{2006}).

\bibitem{Guttler2009}
\bibinfo{author}{Güttler, C.}, \bibinfo{author}{Krause, M.},
  \bibinfo{author}{Geretshauser, R.~J.}, \bibinfo{author}{Speith, R.} \&
  \bibinfo{author}{Blum, J.}
\newblock \bibinfo{title}{{The} {physics} {of} {protoplanetesimal} {dust}
  {agglomerates}. {IV}. {Toward} a {dynamical} {collision} {model}}.
\newblock \emph{\bibinfo{journal}{The Astrophysical Journal}}
  \textbf{\bibinfo{volume}{701}}, \bibinfo{pages}{130--141}
  (\bibinfo{year}{2009}).

\bibitem{Katsuragi_2017}
\bibinfo{author}{Katsuragi, H.} \& \bibinfo{author}{Blum, J.}
\newblock \bibinfo{title}{The physics of protoplanetesimal dust agglomerates.
  {IX}. mechanical properties of dust aggregates probed by a solid-projectile
  impact}.
\newblock \emph{\bibinfo{journal}{The Astrophysical Journal}}
  \textbf{\bibinfo{volume}{851}}, \bibinfo{pages}{23} (\bibinfo{year}{2017}).

\bibitem{Huang2019}
\bibinfo{author}{Rech, F.} \& \bibinfo{author}{Huang, K.}
\newblock \bibinfo{title}{Correction of iq mismatch for a particle tracking
  radar}.
\newblock \emph{\bibinfo{journal}{European Microwave Conference}}
  (\bibinfo{year}{2019}).

\bibitem{Amon2017}
\bibinfo{author}{Amon, A.} \emph{et~al.}
\newblock \bibinfo{title}{Preface: {Focus} on imaging methods in granular
  physics}.
\newblock \emph{\bibinfo{journal}{Rev. Sci. Instrum.}}
  \textbf{\bibinfo{volume}{88}}, \bibinfo{pages}{051701}
  (\bibinfo{year}{2017}).

\bibitem{poschel05a}
\bibinfo{author}{P\"{o}schel, T.} \& \bibinfo{author}{Schwager, T.}
\newblock \emph{\bibinfo{title}{Computational Granular Dynamics}}
  (\bibinfo{publisher}{Springer-Verlag Berlin Heidelberg New York},
  \bibinfo{year}{2005}).

\bibitem{cuda10c}
\bibinfo{author}{Sanders, J.} \& \bibinfo{author}{Kandrot, E.}
\newblock \emph{\bibinfo{title}{CUDA by Example: An Introduction to
  General-Purpose GPU Programming}} (\bibinfo{publisher}{Addison-Wesley
  Professional}, \bibinfo{year}{2010}).

\bibitem{Rubio-Largo2016}
\bibinfo{author}{Rubio-Largo, S.~M.}, \bibinfo{author}{Maza, D.} \&
  \bibinfo{author}{Hidalgo, R.~C.}
\newblock \bibinfo{title}{Large-scale numerical simulations of polydisperse
  particle flow in a silo}.
\newblock \emph{\bibinfo{journal}{Computational Particle Mechanics}}
  \bibinfo{pages}{1--9} (\bibinfo{year}{2016}).

\end{thebibliography}

\section*{Acknowledgements}

We thank Ingo Rehberg, Simeon V\"olkel and Thorsten P\"oschel for helpful discussions. KH and RH acknowledge the support of Bayerische Forschungsallianz through Grant No. BaylntAn-UBT-2018-72 for establishing the collaboration. This work is partly supported by the Deutsche Forschungsgemeinschaft through Grant No.~HU1939/4-1. The Spanish MINECO (FIS2017-84631-P MINECO/AEI/FEDER, UE Projects), supported this work.

\section*{Author contributions statement}

R.H. and K.H. initiated and supervised the research. F.R., V.D. and K.H. performed the experiments. D.D and R.H conducted the simulations. All authors analyzed and contributed to the interpretation of the data. R.H. and K.H. wrote the manuscript. 

\section*{Additional information}

\textbf {Competing financial interests:} The authors declare no competing financial interests.

\end{document}